\documentclass[12pt,preprint]{aastex}
\usepackage{CJK}
\usepackage{lscape}
\usepackage{amsmath}
\usepackage{graphicx}
\usepackage{txfonts}
\usepackage{color}
\usepackage{natbib}
\newcommand{\ltmark}[1]{} % a mark which does not appear
\newcommand{\myemail}{liuteng@ustc.edu.cn}

\newcommand{\name}[1]{{\sl#1}}
\newcommand{\numb}[1]{#1}

\newcommand{\cgs}{erg cm$^{-2}$ s$^{-1}$}

\begin{document}
\begin{CJK*}{UTF8}{gkai}
\slugcomment{Data and images available at \url{http://www.arcetri.astro.it/SWXCS/} and \url{http://swxcs.ustc.edu.cn}}

\title{The Swift X-ray Telescope Cluster Survey III:\\
cluster catalog from 2005-2012 archival data}

\author{Teng Liu (刘腾)\altaffilmark{1}, Paolo Tozzi\altaffilmark{2}, Elena Tundo\altaffilmark{2}, Alberto Moretti\altaffilmark{3}, Piero Rosati\altaffilmark{4}, Jun-Xian Wang (王俊贤)\altaffilmark{1}, Gianpiero Tagliaferri\altaffilmark{5}, Sergio Campana\altaffilmark{5}, Mauro Giavalisco\altaffilmark{6}}

\altaffiltext{1}{CAS Key Laboratory for Research in Galaxies and Cosmology, Department of Astronomy, University of Science and Technology of China, 230026, Hefei, Anhui, P.R. China; \myemail}
\altaffiltext{2}{INAF, Osservatorio Astrofisico di Firenze, Largo Enrico Fermi 5, I-50125, Firenze, Italy}
\altaffiltext{3}{INAF, Osservatorio Astronomico di Brera, Via Brera 28, I-20121 Milano, Italy}
\altaffiltext{4}{Universit\`a degli Studi di Ferrara, Dipartimento di Fisica e Scienze della Terra, Via Saragat 1 I-44121 Ferrara, Italy}
\altaffiltext{5}{INAF, Osservatorio Astronomico di Brera, Via Bianchi 46, I-23807, Merate (LC), Italy}
\altaffiltext{6}{University of Massachusetts, Department of Astronomy, LGRT-B 619E, 710 North Pleasant Street, Amherst, MA (USA)}

\begin{abstract}
We present the catalog of the {\name{Swift}} X-ray Cluster Survey (SWXCS) obtained using the archival data of the X-ray Telescope ({\name{XRT}}) onboard the {\name{Swift}} satellite acquired from February 2005 to November 2012, extending the first release of the SWXCS.
The catalog provides positions, soft fluxes and, when possible, optical counterparts for a flux-limited sample of X-ray group and cluster candidates.  
We consider the fields with Galactic latitude $|b| > 20\degr$ to avoid high $HI$ column densities.
We discard all the observations targeted at groups or clusters of galaxies, as well as particular extragalactic fields not suitable to search for faint extended sources.
We finally select $\sim3000$ useful fields covering a total solid angle of $\sim400$ deg$^2$.
We identify extended source candidates in the soft-band (0.5-2 keV) images of these fields, using the software {\name{EXSdetect}}, which is specifically calibrated on {\name{XRT}} data.
Extensive simulations are used to evaluate contamination and completeness as a function of the source signal, allowing us to minimize the number of spurious detections and to robustly assess the selection function.
Our catalog includes \numb{263} candidate galaxy clusters and groups, down to  a flux limit of $7\times10^{-15}$ \cgs in the soft band, and the logN-logS is in very good agreement with previous deep X-ray surveys.
The final list of sources is cross-correlated with published  optical, X-ray, and Sunyaev-Zeldovich catalogs of clusters.
We find that {\numb{137}} sources have been previously identified as clusters in the literature in independent surveys, while {\numb{126}} are new detections.
At present, we have collected redshift information for {\numb{158}} sources ($60$\% of the entire sample).
Once the optical follow-up and the X-ray spectral analysis of the sources are completed, the SWXCS will provide a large and well-defined catalog of groups and clusters of galaxies to perform statistical studies of cluster properties and tests of cosmological models.
\end{abstract}


% galaxies: high-redshift --

\shortauthors{Liu et al.}

\shorttitle{SWXCS cluster catalog}

\maketitle

\section{INTRODUCTION}
Groups and clusters of galaxies are the most massive, gravitationally bound structures in the
Universe and their hot Intra Cluster Medium (ICM) make them to appear as prominent
extended sources in the X-ray sky.  Therefore, X-ray cluster surveys are among the most 
efficient tools to constrain cosmological parameters and primordial density fluctuations.
A large and complete catalog of groups and galaxy clusters spanning a wide range of redshifts would 
be crucial to make significant steps forward towards the understanding of cosmic structure formation and evolution
\citep{Rosati02,Schuecker05,Voit05,Borgani08}, the chemical and thermodynamical cosmic history 
of the ICM \citep[][]{Ettori04,Balestra07,Maughan08,Anderson09}, and to provide an 
accurate measurement of cosmological parameters \citep[see][]{Vikhlinin09,Mantz10,Allen11}. 

However, this task is not within reach of  current X-ray missions.  All the major X-ray facilities
existing today have not been designed for surveys, and have a low efficiency in 
discovering rare objects like galaxy clusters, particularly at high redshifts.  
The main characteristics required for an effective X-ray survey mission for extended 
sources are: large field of view (FOV, of the order of $1$ deg$^2$), high angular 
resolution (of the order of few arcsec), low background, and a large effective area 
(of the order of $10^4$ cm$^2$).  Looking at the near future, the upcoming mission 
\name{eROSITA} \citep{2010Predehl,2012Merloni} will finally provide an X-ray all-sky coverage 
20 years after the \name{ROSAT} All Sky Survey \citep{1999Voges}, down to limiting fluxes more 
than one order of magnitude deeper than \name{ROSAT} for extended sources.  Therefore \name{eROSITA}
will considerably increase the number of X-ray groups and clusters particularly at
low and moderate redshifts.  However, its limiting flux is predicted to be $\sim 3.4 \times 10^{-14}$ \cgs after four years of operation, 
well above the level below which the majority of the high-$z$ clusters, and medium and high-$z$ groups, are currently found.
In addition, its low effective area above 2 keV severely limits the characterization of the ICM in high temperature clusters \citep[see][]{2014Borm}.

At present, the best resource to build X-ray cluster samples is provided by the still increasing
 archives of the major X-ray facilities, {\sl Chandra} and \name{XMM-Newton}.  For a review of the 
ongoing X-ray cluster surveys, updated to the year 2012, see Table 1 in \citet{Tundo12}.    
% Besides the efforts on detecting new serendipitous clusters from \name{XMM-Newton} and 
% \Chandra archives \citep[][see Table. 1 of \citet{Tundo12}]
% {lloyd-davies11_XCS,fassbender11_XDCP,pacaud06,Clerc12,barkhouse06},
%the \name{XMM} Cluster Survey\citep[XCS,][]{lloyd-davies11_XCS},
%the \name{XMM} Distant Cluster Project \citep[XDCP,][]{fassbender11_XDCP},
%The \name{XMM} Large-Scale Structure survey \citep[LSS,][]{pacaud06},
%the \name{XMM} CLuster Archive Super Survey\citep[X-CLASS,][]{Clerc12},
%and the {\sl Chandra} Multi-wavelength Project \citep[ChaMP,][]{barkhouse06},
In this framework, we recently presented the \name{Swift} X-ray Cluster Survey 
\citep[SWXCS][hereafter Paper I]{Tundo12}, which is based on the archival data of the X-ray telescope 
\citep[XRT,][]{Burrows2005} onboard the \name{Swift} satellite \citep{2004gehrels}.  
In spite of its small collecting area (about 1/5 of {\sl Chandra}) \name{XRT} has two characteristics which 
 make it an efficient instrument for detection and characterization of extended sources: a low 
background \citep{Moretti09} and a 
constant angular resolution (with a Half Energy Width $HEW = 18\arcsec$)  across the entire FOV
\citep{2007Moretti}.  We note that \name{XRT} angular resolution is as good as the resolution  
of \name{XMM-Newton} at the aimpoint, and therefore better than \name{XMM-Newton} when averaged over 
the FOV.  The first catalog of the SWXCS project, including 72 clusters and groups, has been presented in 
Paper I, while the X-ray spectral analysis for more than half of this sample is presented in 
\citet[][hereafter Paper II]{2014Tozzi}.

In Paper I we used only the Gamma Ray Burst (GRB) follow-up observations of \name{XRT} released before April 2010.
The sample we built in Paper I has shown the efficiency of an X-ray telescope as small as \name{XRT} in finding 
and characterizing X-ray extended sources.  
The natural next step is the inclusion of the entire \name{Swift-XRT} archive, which is the goal of this work.
With this aim, we developed a software designed for the detection and photometry of extended sources 
and optimized for the characteristics of \name{XRT} data 
\citep[\name{EXSdetect,}][]{Liu13_EXSdetect}.  The source detection method used in \name{EXSdetect}, is 
a combination of Voronoi Tessellation and friend-of-friend algorithms \citep[VT+FOF,][]{Ebeling93}.
This method does not require {\sl a priori} assumptions about the shape and size of the sources, and it is 
particularly efficient when applied to X-ray images which are characterized by many empty pixels.
Its main limitation consists in the blending effect of merging neighboring sources into one.
Spurious extended sources may occur due to the bridging of two or more faint unresolved sources, or to the 
overlap of true extended emission with the wings of bright, unresolved sources.
To mitigate this effect, in \name{EXSdetect} we developed an accurate deblending procedure which is very effective 
in identifying and separating most of the blended sources, and eventually 
removing the unresolved sources mistakenly included within extended emission.  

The efficiency of our detection algorithm as a function of the exposure time and the source flux, and the accuracy in the source photometry, are investigated by extensive imaging simulations.  
Most importantly, our simulations allow us to evaluate the contamination (number of spurious extended sources) and the completeness of the catalog as a function of the source flux. 
In the simulations we make use of an empirical 
model of the point spread function (PSF) of \name{XRT}.  Thanks to our simulations, we can identify 
an optimal threshold in the source photometry above which our catalog reaches 
the required completeness and purity.  This threshold directly provides a position-dependent 
flux limit for each field, and hence a selection function depending on the physical source flux 
for the entire survey.  This step is particularly relevant since, once the selection function and 
the contamination level are accurately predicted, the sample can be used for statistical studies.  

This work is the extension of the previous SWXCS catalog (Paper I)
to the entire \name{Swift-XRT} archive as of November 2012, which includes more than $10000$ fields, 
as opposed to the $\sim300$ fields used in Paper I.  In addition, we apply for the first time 
the \name{EXSdetect} software to the \name{XRT} data to achieve better accuracy and sensitivity.  
The ultimate goal of SWXCS is to provide  a well-defined, large catalog of X-ray selected 
groups and clusters to investigate X-ray properties and perform classical  cosmological tests.  
The Paper is organized as follows. In \S \ref{chap:buildcatalog}, we describe the selection of the
\name{Swift-XRT}  fields suitable for our serendipitous survey, and  describe the identification 
and classification of the extended sources.  In \S \ref{catalog}, we present the SWXCS catalog 
and compare it with previous works in the field.  In \S \ref{comparison} we compare this catalog
to the previous release.  In \S \ref{correlate} we search for counterparts in the 
optical, X-ray, and Sunyaev-Zeldovich (SZ) catalogs for all our sources.  Finally, our findings are summarized in \S 
\ref{conclusions}.

\section{FIELD SELECTION AND SOURCE IDENTIFICATION\label{chap:buildcatalog}}

\subsection{Field Selection\label{fieldselection}}

From the entire \name{Swift} \name{XRT} archive in the period February 2005 -- November 2012, we select all the fields which can be used to build an unbiased, serendipitous X-ray cluster catalog.
Firstly, we exclude all the fields whose Galactic latitude $\vert b \vert \leqslant20\degr$.
Although these fields could in principle be used to search for bright extended sources, they are typically very crowded, which would cause severe blending problems when spatial resolution is limited.
Moreover, a significant fraction of the soft band emission from groups and clusters would be absorbed by the high $HI$ Galactic column density.
The search for the brightest groups and clusters in the Galactic fields will be performed with a different technique in a dedicated paper (Moretti et al. in preparation).

\begin{figure}[htbp]
\epsscale{0.5}
\centering
\plotone{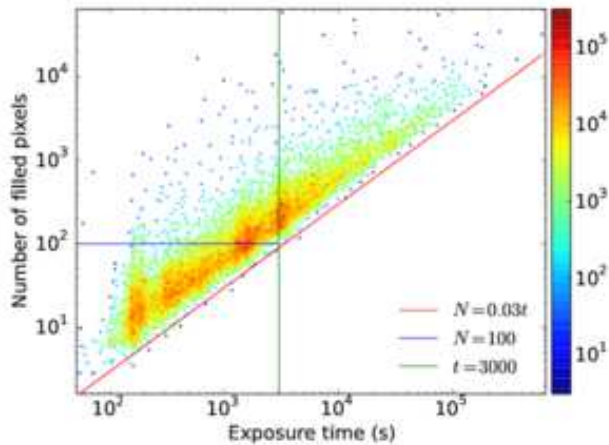}
\caption{Number of filled pixels ($N$) in the soft-band image versus exposure time ($t$) for each field of the \name{XRT} archive as of December 2012.  
The red line ($N=0.03t$) shows the lower envelope of the distribution.
The horizontal blue line shows the condition $N>100$ which ensure a reliable
background measurement, while the vertical black line shows the conservative threshold 
$t_{exp} > 3000$ which guarantees the condition $N>100$.  Note that the large majority
of fields with $t_{exp}< 3000$ and $N>>100$ are dominated by one or few bright sources,
 therefore, despite the large number of filled pixel, their background is poorly characterized.
\label{fig:3ksLowExp}}
\end{figure}

Secondly, we exclude the shallow fields whose exposure time is $t_{exp} \leqslant 3000$ s.
This limit guarantees $\gtrsim100$ total photons in the soft band in each field (see Figure \ref{fig:3ksLowExp}).
This threshold represents the minimum number of photons to sample the background in an X-ray image, 
which is a critical step to identify extended source as enhancement of the photon density 
with respect to the background level.  We set a lower limit of $100$ to the number of total photons based 
on extensive simulations of background-only images, where we tested the capability of 
recovering the true background in our algorithm \citep[see][and Appendix A]{Liu13_EXSdetect}.  This forces us 
to discard a large number of fields (see lines in Figure \ref{fig:3ksLowExp}), which, however, 
would have contributed only at very high fluxes ($\gtrsim10^{-12}$ \cgs), where the expected number of clusters 
is very low. 

We are also forced to exclude the deepest field (164440+573434) whose exposure is 1.9 Ms. 
This choice, which appears to be particularly unfortunate, is due to the large background in the final image. 
While the internal parameters of \name{EXSdetect} are optimized for almost the entire range of exposure
among our fields, \name{EXSdetect} becomes unstable when applied to XRT images with exposure times
around 2 Ms.  Therefore, we choose to discard the field 164440+573434, which would require a different 
approach with respect to the rest of the survey.   We note that the performance of \name{EXSdetect} in the 
second deepest field, whose exposure time is 1.1 Ms, has been successfully  tested with our simulations.

A further,  fundamental step is to filter out all the fields which are directly targeting groups or 
clusters of galaxies, since they would clearly introduce a positive bias towards the detection of extended sources.
In principle, one can simply excise the targeted group or cluster, and use the rest of the field.
However, due to the correlation function of dark matter halos, to have an unbiased sample one should also 
exclude extended sources with similar redshifts.
Since this information is not available for many of our sources, we decide to remove all the fields which, on the 
basis of the target name and coordinates, are aiming at groups and clusters of galaxies.
We also filter out observations targeting nearby galaxies, because such galaxies appear as bright extended 
X-ray sources, whose emission is not associated to ICM but mostly to X-ray binaries and massive 
star formation events.  On the other hand, all the fields targeting at GRBs and AGNs are included in this survey.
In Paper I we already showed that GRBs show no spatial correlation with galaxy clusters, 
neither are AGNs expected to be correlated with clusters.  It is actually expected that AGNs
are suppressed in cluster environments, at least locally \citep{Khabiboulline14}.  Other
multiwavelength studies on the occurrence of AGNs in clusters provide inconclusive results 
\citep[for example,][]{Pimbblet13,Koulouridis14,Ehlert13,Klesman14}.  Recently, it has been found that
the fraction of luminous AGNs in clusters reaches that in the field in the redshift range $1<z<1.5$ \citep{Martini13}.
Therefore, we conclude that no significant bias in cluster detection is expected from 
the inclusion of all the fields targeting at AGNs.
%Actually, AGN activity has been recently found to decrease in cluster environments \citep{Khabiboulline14}.
%No correlation is also expected between groups, clusters and AGN.  In fact, the population of AGN typically found in clusters and groups shows low luminosities \citep[with rare exceptions, like the recently discovered Phoenix cluster,][]{2012McDonald}. 

The task described above is not straightforward, since the target information in the header of \name{Swift-XRT} 
event files is often different from the standard naming conventions or incomplete. 
We go through the keywords of target names and coordinates in all the fields, and identify the targets in the 
NASA Extragalactic Database (NED) when needed.  
Here we describe the selection rules applied to the \name{Swift-XRT} archive in order to identify an unbiased 
subset of fields in details.
Only fields which survived the first triage, i.e., with Galactic latitude $\vert b \vert \leqslant20\degr$ and  
$t_{exp} \geqslant 3000$ s, are considered here.  For completeness we also list the excluded fields.

\paragraph{Selected fields:}
\begin{description}
\item[486] %[001]
GRB follow-up fields, including all the fields previously used in Paper I;
\item[698] %[003]
fields whose targets are found in the NED and are classified as AGN;
\item[654] %[004]
fields whose targets match an AGN within 5\arcsec of their coordinates in the NED;
\item[22] %[005]
\name{Swift-BAT} triggered observations, whose targets are variable hard-X-ray sources;
\item[136] %[010]
fields targeting Fermi/LAT gamma-ray sources (corresponding to header 
keywords: 0FGL, 1FGL and 2FGL \citealp{1FGL,2FGL});
\item[71] %[011]
fields targeting \name{INTEGRAL} gamma-ray sources \citep[header keyword IGR,][]{IGR};
\item[84] %[012]
fields targeting \name{ROSAT} detected sources which are not classified as galaxy clusters in NED; 
\item[401] %[013]
fields whose target names are found in the NED, and are not classified as galaxy clusters or groups;
\item[422] %[014]
fields targeting \name{Swift-BAT} detected sources (mostly local AGN);
\item[10] %[015]
safe pointings, which are carried out when the telescope looses coordinates;
\item[12] %[016]
fields targeting pulsars.
\end{description}

\paragraph{Excluded fields:}
\begin{description}
\item[55] %[101]
fields targeting Abell clusters, Hydra cluster, Coma and the Crab Nebula;
\item[27] %[102]
fields targeting the following nearby galaxies: M31, M33, M63, M67, M81, M82, M87, M100;
\item[17] %[103]
fields targeting comets, which may show diffuse X-ray emission;
\item[157] %[104]
fields targeting nearby galaxies cataloged as NGC or Mrk;
\item[142] %[105]
fields within 11\degr of the Large Magellanic Cloud (LMC) or 6\degr of the 
Small Magellanic Cloud (SMC);
\item[111] %[106]
fields targeting supernovae, which are often hosted by nearby galaxies;
\item[26] %[110]
fields targeting Hickson Compact Groups \citep[HCG,][]{HCG};
\item[2] %[111]
fields targeting \name{ROSAT} detected sources which are classified as galaxy clusters in the NED;
\item[30] %[112]
fields whose target names are found in NED, and are classified as galaxy clusters or groups;
\item[9] %[113]
fields targeting Redshift Survey Compact Groups \citep[RSCG,][]{RSCG};
\item[52] %[120]
fields targeting other sources related to clusters in dedicated observational programs;
\item[8]
fields significantly affected by stray-light;
\item[762] %[0]
fields with unknown target classification.
\end{description}

\begin{figure}[htbp]
\centering
\epsscale{0.8}
\plotone{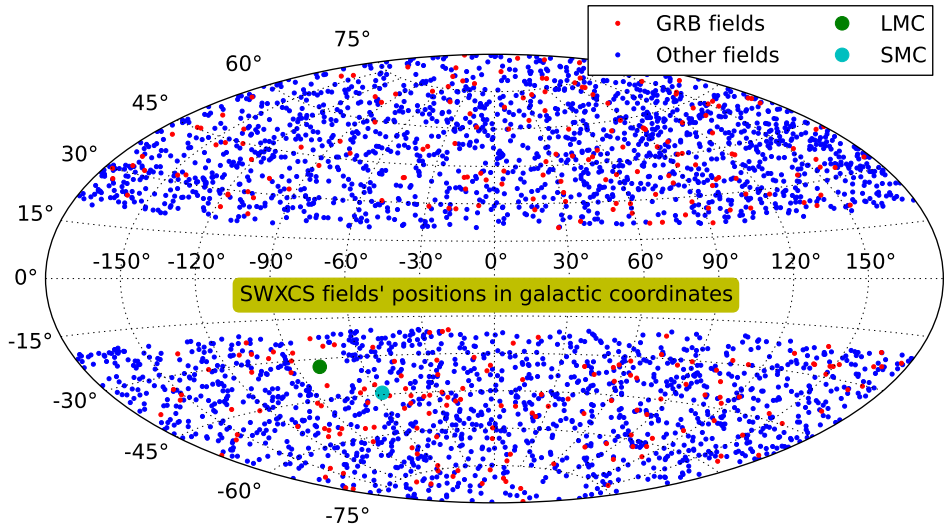}
\caption{Positions of the \numb{3004} selected fields of the SWXCS.  Red points show the GRB fields used in 
Paper I, while blue points show the new fields used in this work.  
The positions of LMC and SMC are marked with large dots.}
\label{fields_positions}
\end{figure}

As previously mentioned, we still have a large number of fields (more than $700$) targeting at sources whose classification is uncertain.
We conservatively discard all these fields with unknown classification, although many useful fields may be lost with this choice.
However, this enables us to avoid any field selection bias, which is a critical requirement for statistical 
studies and cosmological tests.
For consistencies with the first SWXCS catalog, we include 8 fields which would be excluded with these selection rules (because inside the LMC region) but have been used in Paper I.
This inclusion does not have any significant effect on the final catalog.
Finally,  we select \numb{3004} fields which provide a truly serendipitous sampling of the extragalactic sky.  The
positions in the sky of the aimpoints of the selected fields are shown in Figure \ref{fields_positions}.

\subsection{Selection Of Group And Cluster Candidates\label{simulation}}
 
The \name{XRT} data reduction is described in Paper I.  We consider only the soft-band (0.5-2 keV) images for source detection, since the inclusion of the \name{Swift-XRT} hard-band images is not 
useful to identify nor to characterize the detected sources.

At variance with Paper I, where the sources were identified on the basis of a growth-curve method, we use here for the first time the \name{EXSdetect} software.  
This algorithm and its performance on the \name{Swift-XRT} data are described in full details in \citet{Liu13_EXSdetect}.  However, since the software, which is publicly available and currently updated on the 
SWXCS website (\url{http://www.arcetri.astro.it/SWXCS/} and \url{http://swxcs.ustc.edu.cn}), evolved significantly in the meanwhile, in the Appendix we describe in detail the most relevant changes in the 
current version (v3.0).  
The first important change concerns the evaluation of the background  (see Appendix \ref{apd:bkgestimation}), while the second  concerns the source classification scheme (see Appendix 
\ref{apd:classification}).   Another modification introduced in this version, is a different treatment of the sources at large off-axis angles ($\theta >9\arcmin$).
This is necessary since the in-flight calibration of the \name{Swift-XRT} PSF shows that at $\theta \sim 10\arcmin$ the HEW increases significantly by $\sim 40\%$ \citep{2005Moretti}.  
Therefore we run \name{EXSdetect} with a different PSF model to match the expected behavior of the PSF at large off-axis angles, only for the source candidates  at $\theta >9\arcmin$.

We achieve a strong control on the purity and the completeness of the sample thanks to extensive simulations. 
Bear in mind that the simulations are run on a set of synthetic images with same exposure time distribution and same background of the selected SWXCS field. 
Another important aspect is that the input flux distribution of the simulated sources are taken from real, deep data.
Point sources are randomly extracted from a distribution modeled on the number counts found in deep Chandra fields \citep{2002RosatiCDFS,Moretti03,2011Xue,2012Lehmer} and 
simulated down to a flux about one order of magnitudes lower than the expected detection limit of the SWXCS.   
The flux distribution of the input extended sources was taken from the number counts of  groups and clusters measured in the ROSAT deep cluster survey \citep{Rosati98}.
Finally, to take into account the different morphologies of extended sources, we modeled the surface brightness of our simulated sources on real images of ten bright groups and clusters 
of galaxies observed with \name{Chandra}, covering a wide range in ICM temperature \citep[for details see][]{Liu13_EXSdetect}.
We simulate ten times the entire SWXCS survey, which correspond to $\sim 3 \times 10^4$ X-ray images.  

\begin{figure}[htbp]
\centering
\epsscale{0.5}
\plotone{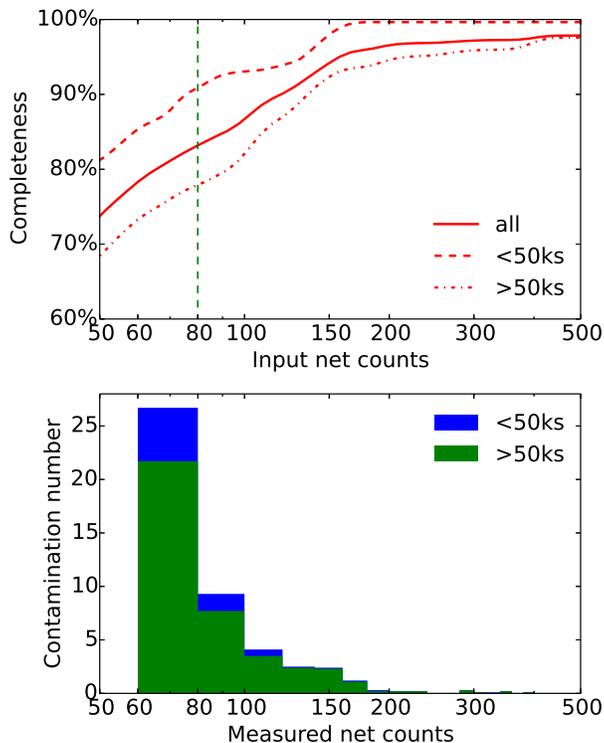}
\caption{{Upper panel:} the solid red line shows the completeness of SWXCS (defined as the ratio of recovered sources and the number input sources) 
as a function of the input net counts as measured with  our simulations. The dashed red lines indicate the completeness obtained only considering fields with an exposure less than 50 ks
(long dashed line) and larger than 50 ks (dot dashed line). 
 {\sl Lower panel:} histogram of the number of spurious sources expected in the entire SWXCS catalog from our simulations in bins of net detected counts.  The different colors show the number of spurious
sources found in fields with exposure time above (green shaded are) and below (blue shaded area) 50 ks.}
\label{fig:sim_com_con}
\end{figure}

In Figure \ref{fig:sim_com_con}, we show the expected completeness (fraction of extended sources recovered at a given value of input net counts), and the expected 
number of spurious extended sources expected in the entire SWXCS in bins of  net detected counts, in the upper and lower  panel, respectively.
We also plot the completeness and contamination separately for fields with exposure time above and below 50ks, which account for 10\% and 90\% of all the fields respectively.
We note that most of the incompleteness and most of the spurious sources come from the 10\% deepest fields.
The reason is the less efficient performance of \name{EXSdetect} in presence of high background and crowded fields.  
For the entire SWXCS, the completeness falls below 90\% at about $130$ net counts, and reaches 83\% at 80 net counts.
Meanwhile, the contamination number increases rapidly below 80 net counts.
Above 80 net counts, the total number of spurious extended sources in the entire survey, at any flux, is estimated to be about $20$, most of which with less than $150$ net counts.   
Therefore, although the completeness is still high and robustly measured below 80 counts, we conservatively set the detection threshold of our catalog to 80 net counts 
within the source region as defined by the \name{EXSdetect} algorithm, to keep a low number of spurious detections.

\begin{figure}[htbp]
\centering
\plotone{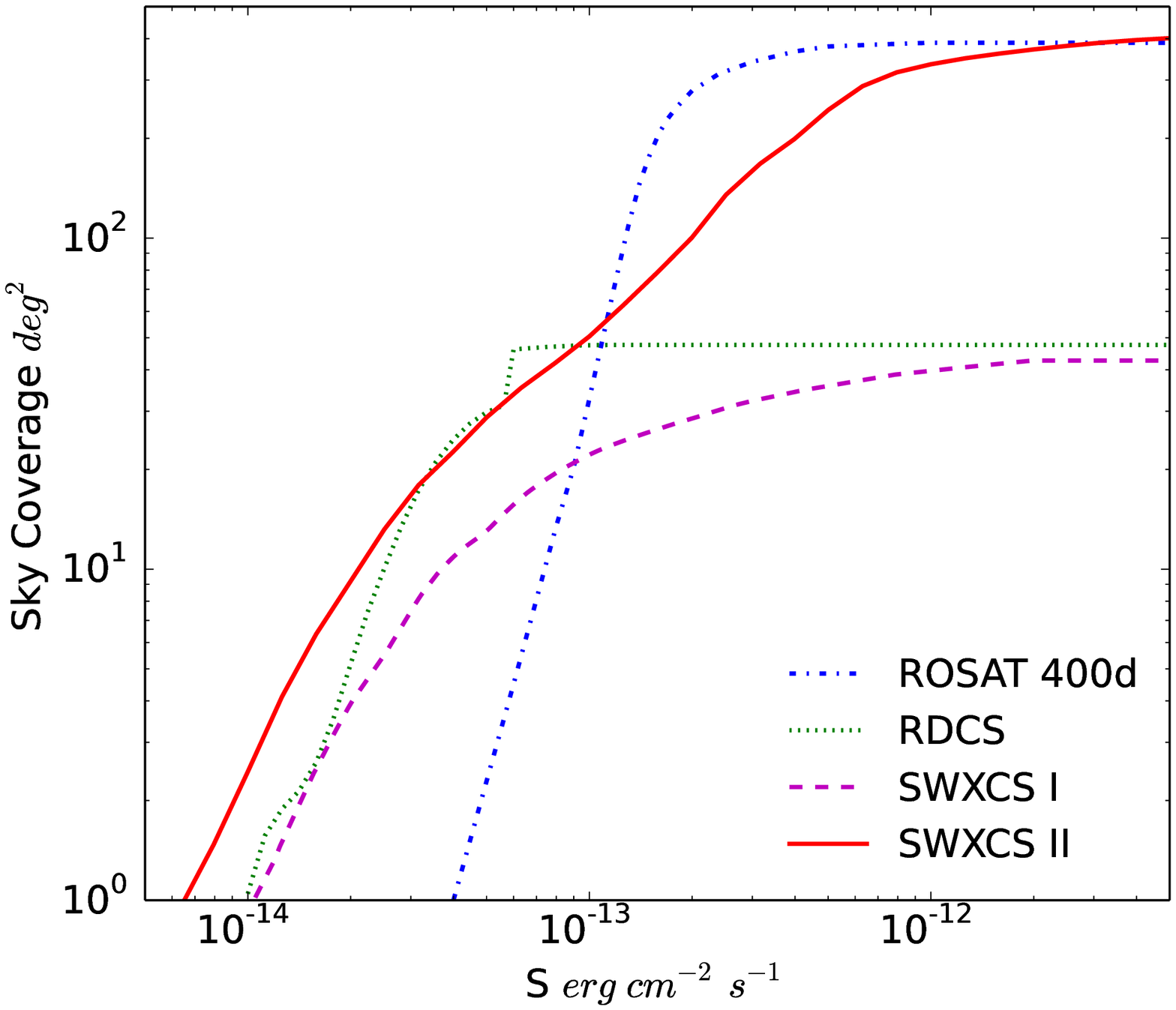}
\caption{Sky coverage of the SWXCS as a function of the soft band flux (solid line).  
For comparison, we also show the sky coverage of the first release of the SWXCS (Paper I, dashed line), of the \name{ROSAT} 400d catalog \citep[][dot-dashed line]{burenin07_400d} 
and of the RDCS \citep[][dotted line]{Rosati98}. }
\label{skycoverage}
\end{figure}

For each field, the minimum detectable count rate, computed as $80/t_{exp}$, corresponds to a position-dependent flux limit obtained by multiplying this number by the field 
energy conversion factor ($ECF$)  at the aimpoint in 0.5-2 keV band (which accounts for the Galactic absorption and is computed for and average thermal model with a temperature of 5 keV, a 
metal abundance of $0.3\ Z_{\odot}$, and a redshift $z=0.4$), and by the normalized exposure map, which accounts for the vignetting effects.  
As shown in Paper I, the $ECF$s depend weakly on the spectral parameters.  Therefore, a flux-limit map is obtained for each field.
The sum of the flux-limit maps of the entire set of fields considered in the SWXCS provides the sky coverage of the survey as a function of energy flux (see Paper I).
The sky coverage of the SWXCS is shown in Figure \ref{skycoverage} (solid line).
The difference with respect to Paper I (dashed line) at low fluxes is mostly due to the lower threshold used ($80$ net counts as opposed to $100$) and to the fact 
that several fields are now deeper thanks to new observations targeting fields already included in Paper I.
We define the flux limit of the survey $S_{lim}$ the flux at which the sky coverage falls below 1 deg$^2$.
This corresponds to a flux of $S_{lim}=7 \times 10^{-15}$ \cgs, which is fainter than that in Paper I ($\sim 10^{-14}$ \cgs).
At the bright end, the increase by a factor of 10 is due to the inclusion of the many shallow fields (mostly not associated to GRB), which were not considered in Paper I.
The maximum solid angle covered by the survey is $\sim 400$ deg$^2$, reached above a flux $S \sim 3\times10^{-12}$ \cgs.
If we compare the sky coverage of the SWXCS with previous deep X-ray surveys of galaxy clusters,  we find that the SWXCS reaches a depth similar to the \name{ROSAT} 
Deep Cluster Survey  \citep{Rosati98} and a width similar to the \name{ROSAT} 400d survey \citep{burenin07_400d} as shown in Figure \ref{skycoverage} (dotted and dot-dashed lines).
Since the sky coverage is essential to derive the number counts and eventually, once the redshifts are available, the source number density as a function of redshift, we provide 
the tabulated values in the second column of Table  \ref{tab:correction}.

\begin{table}[htbp]
\tablenum{1}
\begin{center}
\caption{Tabulated values of the sky coverage and of the completeness as a function of the measured energy flux in SWXCS.  }
\label{tab:correction}
\begin{tabular}{lcl}
\hline
S&Sky Coverage&Ratio\\
\cgs & deg$^2$ &\\
\hline
3.25e-15 &0.1 &42.5\%\\
5.15e-15 &0.6 &50.5\%\\
8.15e-15 &1.6 &60.4\%\\
1.29e-14 &4.4 &72.0\%\\
2.05e-14 &9.7 &80.5\%\\
3.25e-14 &18.5 &85.6\%\\
5.15e-14 &29.6 &90.5\%\\
8.15e-14 &42.9 &95.0\%\\
1.29e-13 &64.5 &97.6\%\\
2.05e-13 &104 &99.4\%\\
3.25e-13 &171 &100\%\\
8.15e-13 &319 &100\%\\
2.05e-12 &372 &100\%\\
8.15e-12 &409 &100\%\\
\hline
\end{tabular}
\end{center}
\end{table}

\begin{figure}[htbp]
\centering
\epsscale{0.6}
\plotone{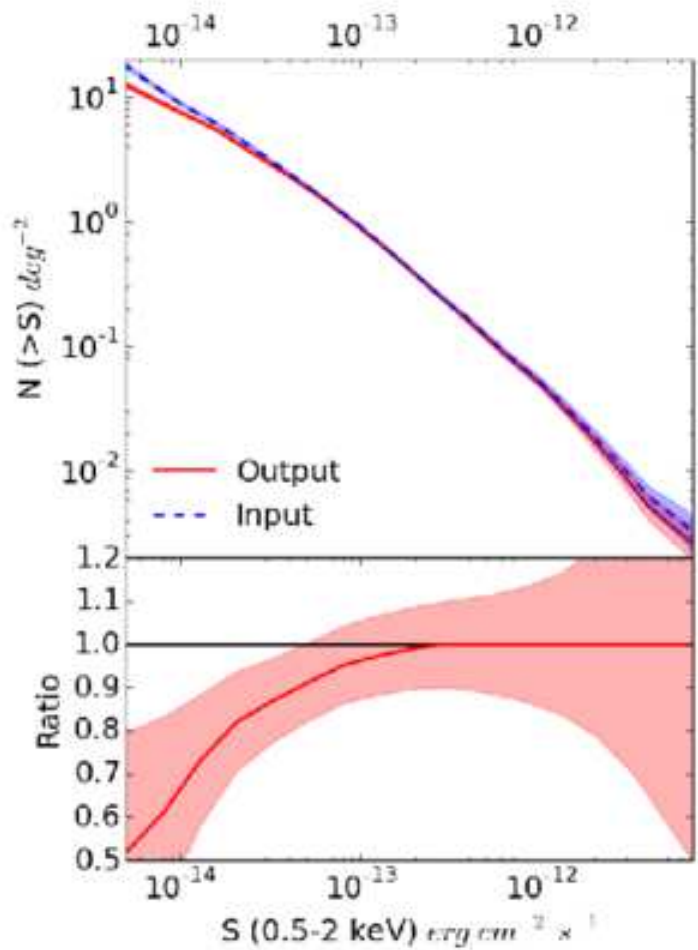}
\caption{{\sl Upper panel}: the blue dashed line and red solid line show respectively the input number counts and the measured number counts from the simulation as a function of the energy flux.
Here we used the entire set of simulations (corresponding to ten times the entire SWXCS survey).
Shaded areas show the $1\sigma$ uncertainty on the input and recovered number counts.
{\sl Lower panel}: the ratio of recovered differential number counts to the input differential number counts, with $1\sigma$ uncertainty, as a function of the energy flux.  This curve, along with the
sky coverage,  is used to compute the number counts in the SWXCS.
}
\label{fig:simu_logNlogS}
\end{figure}

The completeness can also be computed as a function of the energy flux, simply by computing the flux of each source which depends on the source counts rate, the $ECF$ in the field  and 
the and the actual effective area in the extraction region of the source (see Section 3 for details).  The completeness as a function of the flux is then simply obtained as in the previous case, but computing 
the ratio of the recovered over the input sources in bins of energy flux.  In practice, this is equivalent in convolving the completeness function shown in the upper panel of Figure \ref{fig:sim_com_con} with 
the actual distribution of exposure time, $ECF$ and effective area of the SWXCS as represented in the simulations.  In Figure \ref{fig:simu_logNlogS}, upper panel,
we compare the simulated input extended source distribution with the measured extended source distribution recovered with \name{EXSdetect} as a function of the energy flux.  
Note that the input distribution is given by the average of ten actual realizations of the input model used in the simulations, therefore it has a $1 \sigma$ statistical uncertainty shown by the shaded area.
Note also that here the sky coverage is already accounted for.
The curve in the lower panel of Figure \ref{fig:simu_logNlogS} shows the ratio of output to input source distributions as a function of the measured flux.  This function is our best estimate of the completeness of the SWXCS as a function of the energy flux.  We find that the completeness correction is relevant below $5\times10^{-14}$ \cgs, while it is negligible above $10^{-13}$ \cgs.
We remark that, since the completeness is computed as a function of the measured energy flux (and not the actual input flux used in the simulation), this correction takes account 
also the effect of the Eddington bias expected in the SWXCS.
This function is also tabulated in Table \ref{tab:correction}, third column.  The combination of the sky coverage and of the completeness function as a function of the energy flux allows one 
to derive the SWXCS number counts directly from the catalog (see Section 3). 

% We smooth the curve with dynamical scales in order to keep it increasing monotonously with flux.

\subsection{Filtering Of Spurious Sources Not Included In Simulations\label{filterspurious}}

By running \name{EXSdetect} (version 3.0) on the soft band images of the $3000$ fields 
we obtain \numb{430}
% $437$ sources without the removal of the 7 sources at theta~10 arcmin
extended source candidates with a soft band photometry larger than $80$ net counts in the source region defined by the \name{EXSdetect} algorithm itself.
According to the expected performance of \name{EXSdetect}, the source catalog obtained directly should have only about $20$ spurious sources.
Apart from this budget, however, we know that we have spurious sources associated to circumstances not included in the simulations, but that can be easily identified and filtered out.

The first class of these sources are the piled-up targets, which we can always safely assume to be bright AGN or GRB.
At the \name{Swift-XRT} angular resolution, in fact, it is very unlikely that a bright cool core can be affected by pile-up.
Clearly, the pile-up effect modifies the shape of the inner regions of unresolved sources, causing the failure of our source classification algorithm based on the comparison of the source profile with the synthetic image of an unresolved source in the same position and with the same flux.  
For obvious reasons, we did not attempt to include this effect in our simulation.
On the other hand, piled-up sources can be easily identified and removed.  
As a simple and effective criterion to identify piled-up sources, we compute the count rate of each source.
All the data we used are observed in the Photon Counting mode, which has a time resolution of 2.5 seconds \citep{Burrows2005}.
According to the \name{Swift-XRT} documents, any source brighter than 0.5 counts/s in the 0.2-10 keV full band should be checked for pile-up \footnote{see \url{http://www.swift.ac.uk/analysis/xrt/pileup.php}}.
Taking the spectrum of 3C 273 as that of a typical unresolved source, we find that the 0.5-2 keV band contains $\sim 50\%$ of all photons in the full band.
Thus we flag all the sources with 0.5-2 keV count rates $>0.25$ counts/s inside the extraction region as piled-up candidates.
Then, we check their classifications from NED, finding mostly QSOs or galaxies, while none of them, as expected, is associated with clusters or groups.
This step allows us to reject \numb{34} extended source candidates as piled-up sources.  
In addition, some of the GRB may escape this filter since they can suffer pile-up only in high flux states.  Since the identification of all the GRB is straightforward, we filter out \numb{18} GRBs which were mistakenly classified as extended sources.  A simple visual inspection is performed to check whether we may find truly extended sources overlapping the GRB positions, but we found none.

Nearby ($z<0.05$) galaxies constitute another source of contamination for our sample.    
At low redshift, spiral galaxies appear as extended X-ray sources in \name{Swift-XRT} images, which 
represent the populations of high-mass X-ray binaries (in the case of recent starburst) or low-mass X-ray binaries 
in the galaxies. Nearby elliptical galaxies may also show X-ray emission which is related to hot gas in their 
halo.    To identify them we consider the following galaxy catalogs: the  Local Volume Legacy Survey 
\citep[LVL,][]{LVL}, the GALEX Ultraviolet Atlas of Nearby Galaxies \citep{GALEX}, and the third Reference 
Catalog of bright Galaxies \citep{RC3}, and select all the galaxies whose major axis diameter is larger than 
the half power diameter of the \name{XRT} PSF ($18\arcsec$).  For these sources, the extended X-ray 
emission strongly overlaps with the optical extent of the galaxies, as seen in the Digital Sky Survey (DSS) optical 
images or the Sloan Digital Sky Survey (SDSS) images when available, and therefore is clearly dominated by stellar 
sources.  In some cases, we notice some emission beyond the optical extent of the galaxies.  This is particularly 
evident in galaxy pairs. However, in all the cases where the extended emission may be associated with hot gas 
around the galaxies, its emission is weak and  very hard to decouple from the stronger X-ray emission from the 
disk or the bulge.  In other cases, some small-extent diffuse emission is swamped by a central AGN.  For all 
these sources an accurate modeling of the non-thermal X-ray emission is needed before the thermal 
component can be properly evaluated.  Given the very small impact that such contributions would bring to our 
final catalog, all the \numb{30} extended source candidates associated with nearby galaxies are excluded. 

Another kind of spurious detections are caused by bright optical sources (generally stars) whose intense optical/UV emission induces significant spurious charge load in the CCD.  Such sources are 
automatically screened by \name{Swift-XRT} pipeline.  However, this process often leaves a ring-like signature which is classified as extended by our algorithm, but can be immediately spotted by visual 
inspection. We find and reject \numb{10} of these spurious detections.

Another effect that, for obvious reasons, is not considered in the simulations, is associated to the wings of bright clusters which only partially fall in the FOV of the \name{XRT} image.
By visually inspecting RASS, or \name{XMM-Newton} images when available, centered on the source position, we identify \numb{11} cases in which the wings of large X-ray clusters have been detected 
as extended source.
The classification of these sources by \name{EXSdetect} is indeed correct, however we will not include them in the catalog, since the majority of the cluster emission is, in all the cases, well beyond the \name{XRT} FOV. 
The clusters responsible of these detections are Coma, Abell 0496, Abell 1285, Abell 1387, Abell 1767, Abell 2199, Abell 2199, Abell 2877b, Abell 3334, Abell 3395 SW, and MCXC J1423.8+4015. 

The X-ray images sometimes can be contaminated directly by  sunlight, which creates a flare in the light curve and significant diffuse emission at the image borders.  By checking the light curve 
of extended sources at the image borders, we find that the emission of \numb{22} source candidates is actually due to optical flares. These sources are discarded as optical contaminations.

We also find that in \numb{11} cases our extended source candidates are associated with the position of the nominal target of the \name{Swift-XRT} observation within a distance of $2\arcmin$.
Among these, nine are bright \name{ROSAT}   \, X-ray sources, which makes them possible galaxy clusters detected but not identified by \name{ROSAT} because of its poor spatial resolution.
In other words, although the targets of \name{ROSAT} sources were not identified as clusters (see \S \ref{fieldselection}), they have higher probabilities to be clusters than random objects.
The other two targets are galaxies which are not associated with known clusters.
Apparently, these last two sources may be allowed in a serendipitous sample.
However, the perfect position match ($<6\arcsec$) indicates a strong connection between the galaxy targeted by \name{Swift-XRT} and the X-ray extended source.
Also in these two cases, a positive bias is introduced because the possibility of finding a cluster associated to the targeted galaxy.
Overall, although these \numb{11} targeted observations survived our field selection (\S \ref{fieldselection}), they may have higher probabilities to host groups or clusters 
with respect to truly serendipitous observations.  Therefore we remove these \numb{11} extended source candidates from the final list.

Finally, we consider a last case which is not properly treated in our simulations.
We randomly sampled the number counts of unresolved sources on a solid angle of $400$ deg$^2$.
This is what actually happens for genuinely random fields.
However, a significant fraction of fields in the \name{Swift-XRT} archive are targeting very bright QSO.
In some of these cases we find anomalous extended sources which likely to be due to spurious effects 
associated to the X-ray emission of the bright QSO.  In particular, in two fields targeting extremely bright quasars, 
we find $6$ sources in the outskirts which are most likely
associated to the anomalous background due to the presence of the bright source. %072151+712118 225353+161055
We also find $5$ very bright sources which are surrounded by a much fainter extended emission.  
In these few cases, the extended emission component is sufficient to classify these sources
as extended, however the central source is not identified as unresolved, probably due to the
contribution of the extended emission, so it is not removed, as it is done in all the cases when 
unresolved sources are embedded in extended emission.  These cases should be 
treated separately, and a PSF deconvolution of the unresolved source emission should be 
applied before the extended emission could disentangled from the unresolved emission and properly measured.  Since this procedure would introduce large errors on the photometry of the faint extended emission, we decide to remove these sources from the catalog, despite the fact that they do include extended emission.  Clearly, the angular resolution is a strong limitation which hampers us to properly deal with such cases.  Overall, we remove \numb{11} sources due to
effects associated to the presence of targeted, very bright unresolved sources.

At the end of this cleaning procedure, we have removed \numb{147}
%174 (wc EXS_catalog.knownbad) - 7 9arcmin_off-axis sources - 20 visualinspection sources
sources due to effects which could not be included in the simulations.  In principle, these effects could
be implemented in the reduction pipeline with some additional effort.  However, in our case, 
due to the limited number of sources in the SWXCS, a manual check {\sl a posteriori} is feasible
and the automatization of this filtering process is not crucial at tis stage of the project.  
After this step is completed, we are left with \numb{283} group and cluster candidates whose properties are well
described by the completeness and contamination function obtained with our simulations.

\subsection{Beyond The Software: Learned Visual Inspection\label{visualinspection}}

Assuming that we have filtered out any possible source of contamination not included in 
the simulations, we are now dealing with a sample with known statistical properties. 
However, we can extract more information from the simulations.  If we revise all the spurious
sources we find in the simulations (several hundreds) we can understand in most of the cases the
reasons why the Voronoi algorithm failed.  For example, the quite common case of blending of
two or more visual sources can be easily associated to a particular pattern in the surface
brightness distribution of the spurious source candidate.
%In other cases, the change alignment of faint unresolved source result in a strongly elongated, apparently extended source. %NO SUCH CASE
This visual learning procedure is very  effective and in principle could be implemented 
in the software as a machine learning process, as is now commonly done for
data mining in very large surveys.  However, the human eye still appears to be the best tool to 
implement such complex processes.  This has been clearly shown by the “crowd-sourcing” 
projects proposed by  Galaxy Zoo \citep[see][]{2011Lintott} in the last years, which 
have been proved to be highly successful.  It may be very useful, in the future, 
to set up such a crowd-sourcing astronomical project based on X-ray images.

In our case, given the small size of our source list, we are able to train the eye with simulation and
manually apply the filter with a direct visual inspection.  The sources which are flagged as spurious
after this step are \numb{20}, consistent with 
the number of spurious sources expected from the simulations ($\sim 20$).  
Therefore, we decide to include this step in our source classification scheme.   
In the end, we are left with a final catalog containing \numb{263} source candidates of clusters and groups,
whose selection function is shown in Figure \ref{fig:sim_com_con}.  After the visual inspection,
we are confident that the contamination in the SWXCS catalog is reduced at a level which 
can be safely ignored when deriving statistical properties of the sample. 

%10 of them were found to be blending of more than one unresolved sources which were 
% very near to each other, 16 appeared as unresolved sources to the human eye, 
% and the other 9 seemed to be background fluctuations.

\section{SWXCS CATALOG \label{catalog}}

\begin{figure}[htbp]
\epsscale{0.6}
\centering
\plotone{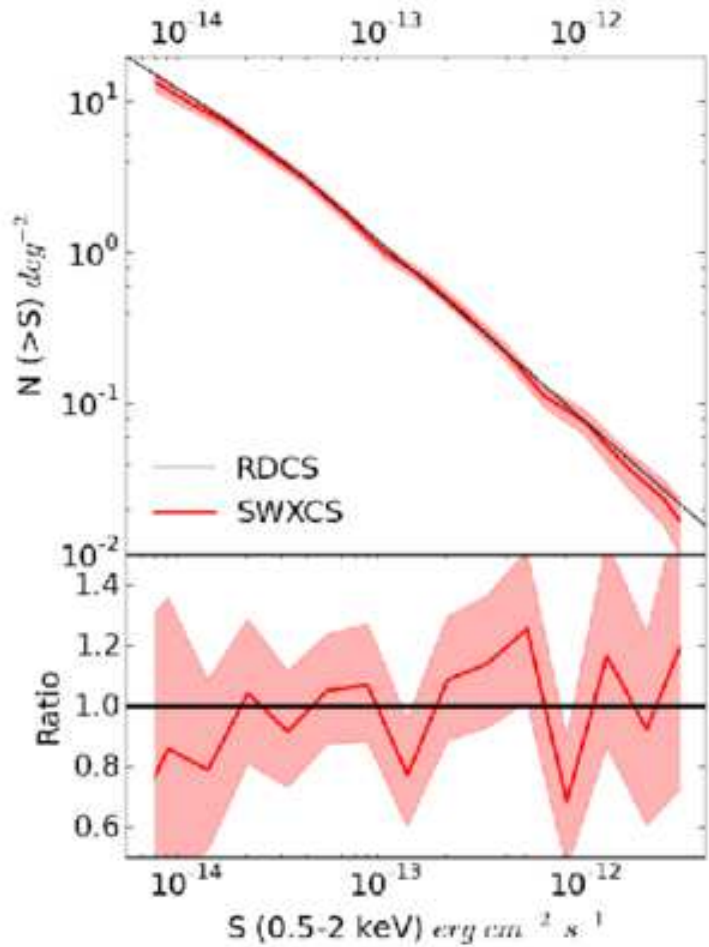}
\caption{{\sl Upper panel:} the red solid line shows the corrected number counts of SWXCS.
%The uncorrected number counts are also plotted (green dashed line), to show the entity of the correction.
The black line is the best-fit double power-law model of RDCS \citep{Rosati98}. 
Shaded areas show the corresponding $1\sigma$ confidence intervals in both panels.  
{\sl Lower panel:} the ratio of SWXCS over RDCS differential number counts.}
%, before (green dashed line) and after (red solid line) being corrected.}
\label{fig:numbercount}
\end{figure}

%GRB follow-up fields have been demonstrated to be a good choice for galaxy cluster survey (Paper I). We compare the logNlogS measured within the GRB follow-up fields as used in Paper I, and the one measured within the other ``good'' fields. As shown in Figure \ref{fig:numbercount_GRBnoGRB}, the identical logN-logSs indicates no selection bias.
%\begin{figure}[htbp]
%\centering
%\includegraphics[width=\columnwidth]{figures/logNlogS_GRBnoGRB.pdf}
%\caption{The logNlogS measured within the GRB follow-up fields as used in Paper I (red) and the one measured within the other ``good'' fields (blue). 68\% confidence intervals are plotted.}
%\label{fig:numbercount_GRBnoGRB}
%\end{figure}

In Table \ref{tab:catalog} we list the \numb{263} sources of the catalog with their X-ray properties.  
The sources already presented in Paper I are marked with an asterisk.  The catalog contains the
following information:

\begin{itemize}
\item Column 1: source name according to the format officially accepted by the IAU Registry in February 2013. 
The format is SWXCS JHHMMSS+DDMM.m.  
This format is different from that used in Paper I, but it was already used in Paper II.
Note that for the sources that presented in Paper I, we keep the same positions used in Paper II, 
although the new positions typically differ by  $\sim 5\arcsec$.
\item Columns 2 and 3: RA and DEC coordinates of the X-ray centroid, defined as weighted median position of the $27$ brightest pixels in the source region (each pixel is weighted
by its density which equals to the pixel value divided by Voronoi cell area).
\item Column 4: the effective exposure time of each source computed as:
\begin{equation}
t_{eff}\ =\ \frac{\sum{n_i}}{\sum{n_i/t_i}} \, ,
\end{equation}
where $i$ is the index of the filled pixels within the source extraction region,
$n_i$ is the photon count in the $i^{th}$ pixel, $t_i$ is the corresponding value in the vignetted exposure map.
\item Column 5: value of the Galactic neutral hydrogen column density in unit of $10^{20}$ cm$^{-2}$, as found in the Leiden/Argentine/Bonn radio survey \citep{2005Kalberla}.
\item Column 6: $R_{eff}$ in arcsec, defined  such that $\pi R_{eff}^2$ equals the source extraction area (which has no {\sl a priori} constraints on its shape).
\item Column 7: the net counts $N_{net}$ measured in the soft band image by \name{EXSdetect} inside the source extraction region found by the VT+FOF method, after background subtraction and removal of unresolved sources.  At present, no correction is applied to compensate for the lost diffuse emission in the region of the removed unresolved source.
The $1\sigma$ error is computed as $\sqrt{N_{tot}+N_{bkg}}$, where $N_{bkg}$ is the counts of background photons, and $N_{tot}\ =\ N_{net}+N_{bkg}$.
\item Column 8: signal-to-noise ratio (S/N) in the soft band, computed as the net counts divided by the associated error.
\item Column 9: estimated soft band flux in units of $10^{-14}$ \cgs.  
To compute the flux, for each field, we estimate the energy conversion factor ($ECF$) in 0.5-2 keV band, taking into account the Galactic absorption in this field, assuming a hot diffuse gas emission model with a temperature of 5 keV, a metal abundance of $0.3\ Z_{\odot}$, and a redshift $z=0.4$.
As shown in Paper I, the $ECF$ depends weakly on the spectral parameters.
The flux of each source is calculated as $S = N_{net} \times ECF / t_{eff}$.  
The $1\sigma$ error of flux is measured by propagation considering the error of net counts and a 4\% systematic error of $ECF$ due to the typical uncertainty in the actual
spectra shape of each source (see Paper I).
More accurate fluxes will be available from the X-ray spectral analysis of our sources, which is postponed to a forthcoming paper (Moretti et al. in preparation).
\end{itemize}

Finally, using the sky coverage and the completeness function described in   \S \ref{simulation}, we compute the number counts of the SWXCS catalog, following the same procedure as described in Paper I.
The corrected number counts are shown in the upper panel of Figure \ref{fig:numbercount}, with $1\sigma$ confidence intervals, which includes the Poissonian error and the uncertainties 
on the average conversion factors.   As we also found in Paper I, the number counts are 
consistent with the logN-logS of the \name{ROSAT} Deep Cluster Survey (RDCS) \citep{Rosati98} (see lower panel of Figure \ref{fig:numbercount}).
The faint end is  also consistent with the very-deep number counts measured from the Extended Chandra Deep Field South (Finoguenov et al. in preparation).
We find that the differential number counts, after correction for incompleteness, can be fit with a broken power-law.  The best-fit model is:
\begin{equation}
{{dN}\over{dS}} = \left\{
\begin{aligned}
&(1.29\pm 0.26) \times \left(\frac{S}{10^{-13}}\right)^{-2.14\pm0.06},\ S>(5.1\pm2.0)\times 10^{-14}\\
&(1.75\pm 0.35) \times \left(\frac{S}{10^{-13}}\right)^{-1.68\pm0.16},\ S<(5.1\pm2.0)\times 10^{-14}\ ,
\end{aligned}
\right.
\end{equation}
where S in unit of \cgs.
Therefore, the slope of the faint end appears to be flatter than the slope at the bright end, although with a low significance($< 3\sigma$).

% OLD FIT TO THE DIFFERENTIAL NUMBER COUNTS, VERSION2.0
%\begin{eqnarray*}
%&& {{dN}\over{dS}} = (1.30\pm 0.28)  \times \Big({{S}\over{10^{-13}}}\Big )^{-2.18}\, \, , S >5\times 10^{-14} \\
%&& {{dN}\over{dS}} = (1.77\pm 0.38 ) \times \Big({{S}\over{10^{-13}}}\Big)^{-1.71}\, \, , S <5\times 10^{-14}\, \, .
%\end{eqnarray*}

\section{\label{comparison}COMPARISON WITH PAPER I}

\begin{figure}[htbp]
\centering
\plotone{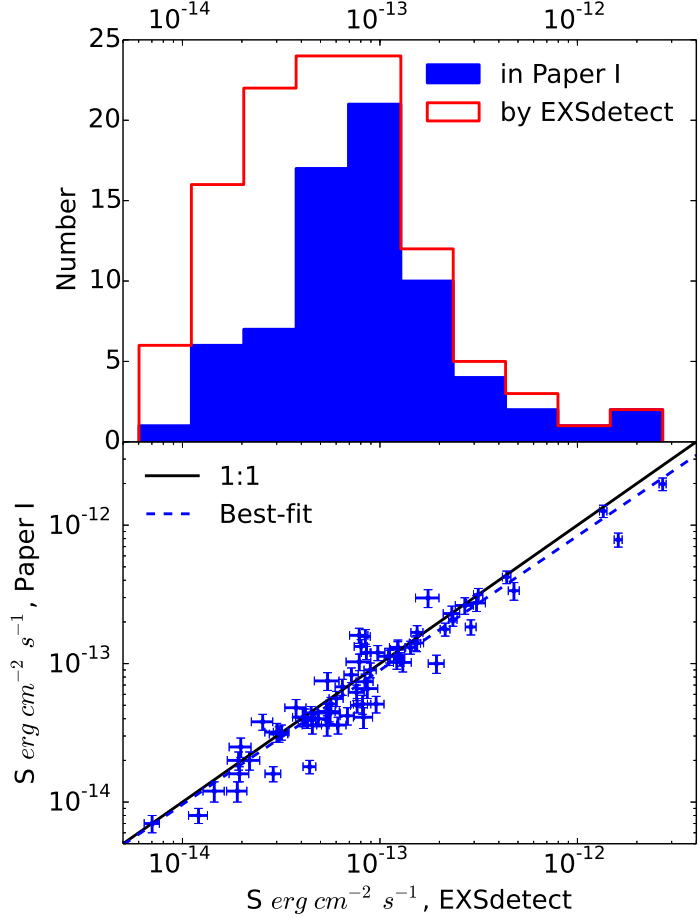}
\caption{{\sl Upper panel}: histogram distributions of the \name{EXSdetect} measured soft-band fluxes of all the sources detected with \name{EXSdetect} in the same GRB fields as used in Paper I (red histogram) and of the \numb{71} SWXCS sources that already presented in Paper I (blue filled histogram).
{\sl Lower panel}: for the \numb{71} SWXCS sources that Paper I has in common, the soft-band fluxes $S_{PaperI}$ measured in Paper I within $R_{ext}$, compared with the soft-band fluxes $S_{EXS}$ measured in this paper with \name{EXSdetect} from a source region defined by the VT+FOF algorithm.
Black solid line shows the relation $S_{PaperI} = S_{EXS}$, while the blue dashed line shows the best fit $\log(S_{PaperI}/10^{-13})\ =\ 0.97 \times \log(S_{EXS}/10^{-13}) -0.05$.}
\label{Compare_CatalogI}
\end{figure}

The first release of the SWXCS catalog (Paper I) was based on a much smaller number of fields ($\sim300$ GRB follow-up fields), and was obtained using a 
standard wavelet detection algorithm coupled with a growth-curve method used to characterize extended sources.
To compare this work with the previous release, first we investigate the effect of using \name{EXSdetect}, which is applied to real data for the first time in this work. 
After filtering out all the spurious sources as described in \S \ref{filterspurious} and \S \ref{visualinspection}, \name{EXSdetect} detects \numb{113} sources in the 
GRB follow-up fields which were used in Paper I (and clearly also included in this work).
All of the 72 sources presented in Paper I except one (SWXCS J022344+3823.2) are recovered by \name{EXSdetect}. In addition, \numb{42} new sources are detected for the first time by \name{EXSdetect}.
Among the newly detected sources, $17$ have less than 100 net counts. 
The other $25$ new sources, instead, have a photometry brighter than 100 net counts, and therefore 
should have been included in the first release of the catalog in Paper I. So we conclude that they 
were simply missed by the detection method used in Paper I.
This shows that the \name{EXSdetect} algorithm is more efficient, allowing us to recover, above the same 
photometry threshold, a number of sources 30\% higher than in Paper I. In the upper panel of 
Figure \ref{Compare_CatalogI}, we plot the flux distribution of the $113$ \name{EXSdetect} detected 
sources, compared with the flux distribution of the sources
of Paper I.  Most of the newly detected sources are found at fluxes below $10^{-13}$ \cgs, showing 
that \name{EXSdetect} is able to reach higher sensitivity allowing
us to further explore the flux range where medium and high-$z$ clusters are found ( $few \times 10^{-14}$ \cgs).

\begin{figure}[htbp]
\centering
\plotone{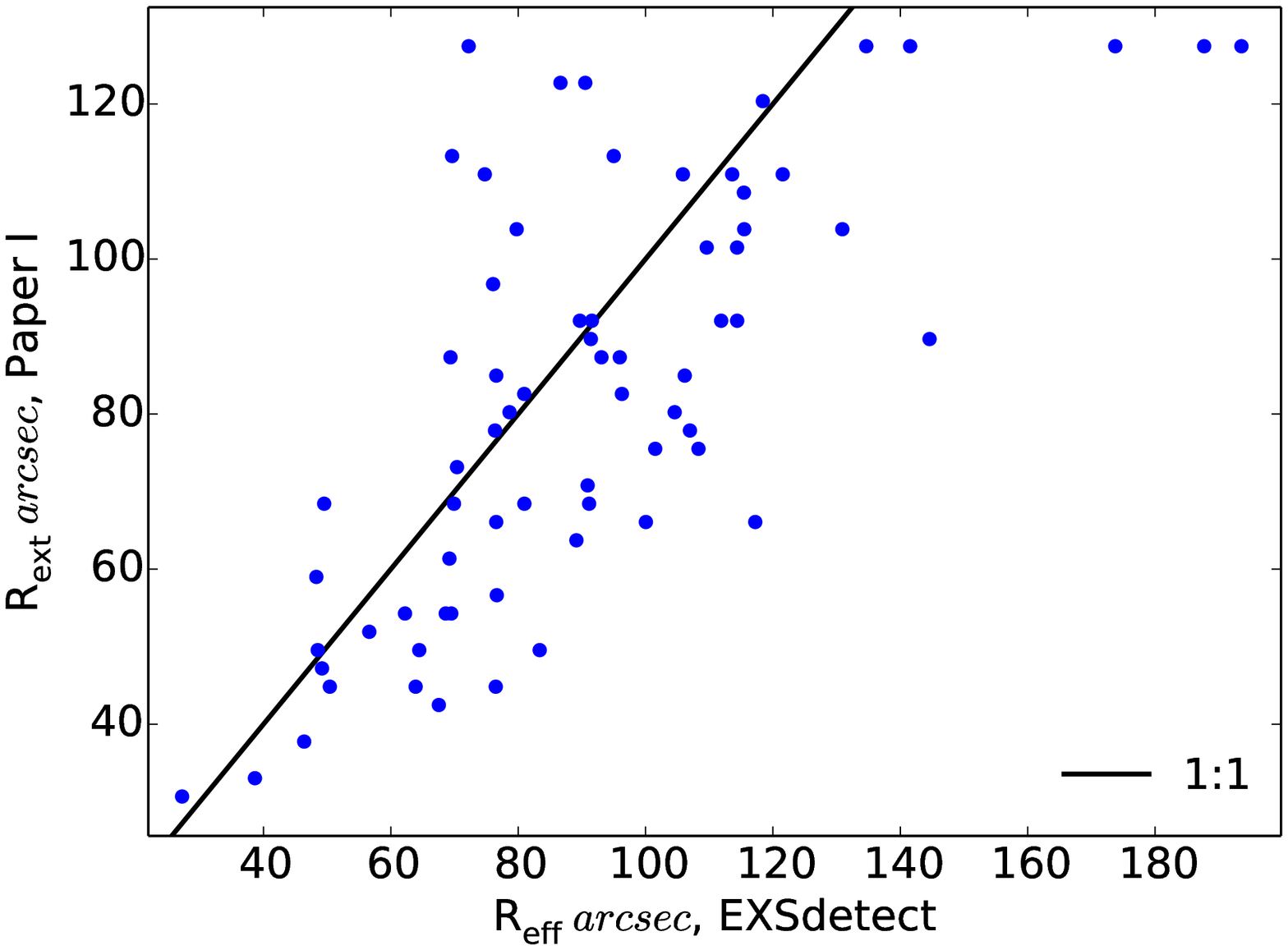}
\caption{Extraction radius $R_{ext}$ used in Paper I to define the source region, 
compared to the effective radius $R_{eff}$ for \name{EXSdetect} sources.  
The solid line shows the relation $R_{ext} = R_{eff}$. }  
\label{fig:Rad_Rad}
\end{figure}

In the lower panel of Figure \ref{Compare_CatalogI}, we show that the photometry obtained with 
\name{EXSdetect} is in good agreement with the values found in Paper I for the \numb{71} sources in common.
Note that the fluxes measured in Paper I are corrected for the missed flux beyond $R_{ext}$ assuming a best-fit $\beta$ model, while in this work no correction is applied.
The best-fit relation between \name{EXSdetect} fluxes and the fluxes in Paper I reads:
\begin{equation}
\log\left(\frac{S_{PaperI}}{10^{-13}}\right)\ =\ (0.97\pm 0.13)  \times \log\left(\frac{S_{EXS}}{10^{-13}}\right) -0.05 \, .
\end{equation}
%This implies that the photometry provided by \name{EXSdetect} is somewhat larger than the photometry measured Paper I, up to a maximum of $\sim 10$\%  for the brightest sources.
The \name{EXSdetect} fluxes are somewhat higher than the fluxes measured in Paper I, up to a maximum of $\sim 10$\%  for the brightest sources.
On the other hand, our simulations showed that the \name{EXSdetect} photometry is accurate at the level of 
$\sim 1-2 $\% \citep[see Figure 10 in][]{Liu13_EXSdetect}.    
We remind the reader that \name{EXSdetect} automatically defined
irregular extraction regions thanks to the Voronoi algorithm, as opposed to the circular extraction regions 
defined in Paper I.  
We find that generally the \name{EXSdetect} extraction region is larger than the circular region used in Paper one, as shown in Figure \ref{fig:Rad_Rad}, by comparing the $R_{eff}$ with the extraction 
radius $R_{ext}$ which was defined in Paper I as the radius where the average source flux equals the background level.
Note that \numb{2} sources whose $R_{eff}$ is more than twice larger than $R_{ext}$ do not appear in this figure, because of large, low surface brightness extents associated to these sources, 
which was not accounted for in Paper I.  Therefore, we conclude that the use of  extraction region defined by the Voronoi method is more efficient in recovering the flux in the low surface brightness outskirts 
of extended sources, providing a more accurate estimate of the total flux.

\begin{figure}[htbp]
\centering
\plotone{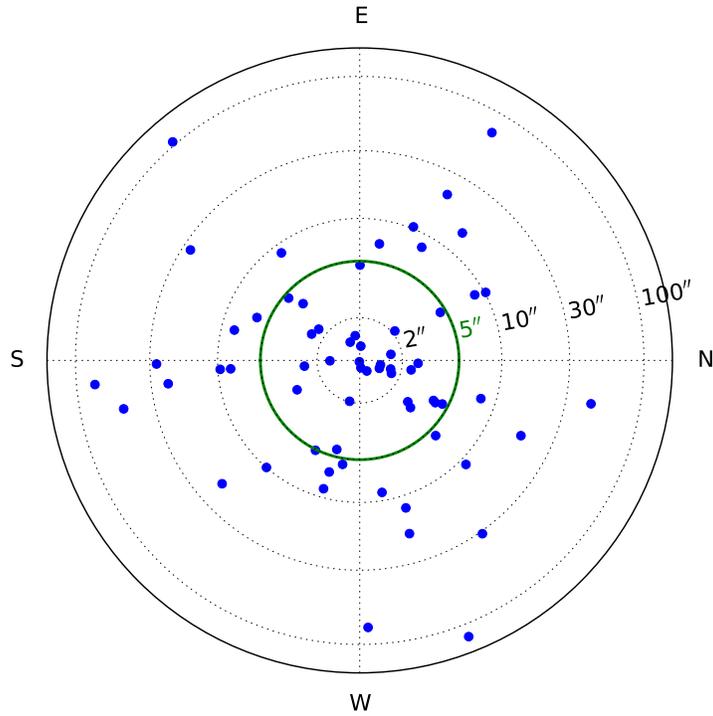}
\caption{Displacements between the source positions in Paper I and the positions found in this work, 
in units of arcsec.  The green circle shows the median separation corresponding to 5\arcsec.}
\label{fig:displacements}
\end{figure}

Finally, for the sources in common, we show the distribution of the displacements between the positions published in Paper I and the positions measured by \name{EXSdetect} (see Figure \ref{fig:displacements}).
The discrepancy is mostly due to the different definition of the center of the source used in this work and, 
to a lesser extent, to the larger extraction regions.
Despite this, the center of the large majority of the sources is changed by less than $20\arcsec$, a value very close to the HEW of the \name{Swift-XRT} PSF.
Eight sources are found at separations between 0.5\arcmin and 2\arcmin, because of the large extent and the rather flat surface brightness distribution of these sources.
Although the change in the nominal position of some source would imply a change in the name according to the IAU format, we prefer to keep the same name used in Paper II for the sources of the first release of the SWXCS catalog.

\section{CROSS-CORRELATION WITH OPTICAL, X-RAY AND SZ CATALOGS AND OPTICAL FOLLOW-UP\label{correlate}}

We checked for counterparts in previous X-ray cluster surveys, in optical cluster surveys, and in the Planck SZ cluster survey.
We simply assume a search radius of $2\arcmin$ from the X-ray centroid, which has been shown to be an efficient criterion  in Paper I.  Nevertheless, we also inspected the area within $5\arcmin$ from the X-ray centroid to investigate whether some possible identification is found at radii larger than $2\arcmin$.
Counterparts at distance between $2\arcmin$ and $5\arcmin$ are included when the optical or SZ corresponding source has a large uncertainty in position.
This is often the case for optical, sparse clusters, or for SZ cluster candidates.
We list all the counterparts associated to the SWXCS sources in Table \ref{tab:matched}, with the measured redshift when available.
In case of multiple counterparts, we list all of them.
Except for a few cases where we have multiple counterparts with statistically inconsistent redshifts, we keep the counterpart with the smallest
distance from the X-ray center.  

From optical surveys, we found \numb{233} optical counterparts corresponding to \numb{116} SWXCS sources, 
including \numb{84} from the SDSS WHL catalog \citep{Wen12}, \numb{25} from the SDSS AMF 
catalog \citep{Szabo11_AMF}, \numb{28} from the SDSS MaxBCG catalog \citep{Koester07_MaxBCG}, 
\numb{45} from the SDSS GMBCG  catalog \citep{GMBCG}, \numb{8} from the SDSSC4 catalog 
\citep{2005SDSSC4,2007SDSSC4}, \numb{27} from the Abell catalog \citep{Abell89}, 
\numb{8} from the NSCS1 catalog \citep{NSCS1}, \numb{4} from the NSCS2 catalog \citep{NSCS1}, \numb{3} from the EDC catalog \citep{1992Lumsden}, and, at last, \numb{1} from the SDSS galaxy groups and clusters catalog built by \citet{Berlind06}.
The majority of the SWXCS sources with optical counterparts are listed in more than one catalogs.  
A few WHL counterparts published in Paper II are found with different names in this work, because of the updated version of WHL catalog used here.

From X-ray surveys, we found \numb{70} X-ray counterparts classified as cluster, corresponding to
\numb{36} SWXCS sources.  In detail, we found \numb{12} X-ray clusters in the \name{ROSAT} 400d 
catalog \citep{burenin07_400d}, \numb{11} in the Northern \name{ROSAT} All-Sky (NORAS) catalog
\citep{Bohringer00_NORAS}, \numb{8} in the \name{ROSAT}-\name{ESO} flux Limited X-ray galaxy cluster catalog \citep[REFLEX][]{Bohringer04_REFLEX}, \numb{3} in the \name{XMM-Newton} Cluster Survey (XCS) catalog  \citep{Mehrtens12_XCS}, and \numb{1} in the \name{Chandra} Multiwavelength Project (ChaMP) galaxy cluster catalog \citep{ChaMP}.  We also found \numb{35} counterparts in the MCXC catalog \citep{MCXC}, which includes most of the X-ray clusters above.

Finally, for \numb{15} SWXCS sources, we found \numb{16} cluster counterparts detected via SZ effect, \numb{13} by Planck \citep{Planck13} and \numb{3} by South Pole Telescope \citep[SPT,][]{SPT}. The Planck sources are typically at larger distances from the X-ray centroid (between $1\arcmin$ and $3\arcmin$), because of the much larger position errors of Planck clusters \citep[see][]{2007Schafer}. 

Overall, about half (\numb{137}) of the \numb{263} SWXCS sources have been previously  identified as groups or clusters of galaxies, 
while \numb{126} SWXCS sources are new cluster and group candidates.  Thanks to these identifications, we are able to recover the redshift information for a significant fraction of our sample.  We collect spectroscopic or photometric redshift for \numb{130} of our sources.  Moreover, to increase the number of available redshifts, we also search in NED catalogs for single galaxies with published redshift not associated to previously known clusters within a search radius of $7\arcsec$ from the X-ray centroid of our sources.  We find \numb{50} galaxies with measured redshift for \numb{47} of our sources, as a complement to the redshifts obtained from cluster counterparts.
In \numb{35} cases where we have both cluster and galaxy counterparts, the galaxy redshifts are consistent with those of clusters.
% the single galaxy is simply the central galaxy of a cluster counterpart we already found.
In the \numb{12} cases where no cluster counterpart is found, we tentatively assign the galaxy redshift to our X-ray source.  

\begin{figure}[htbp]
\centering
\plotone{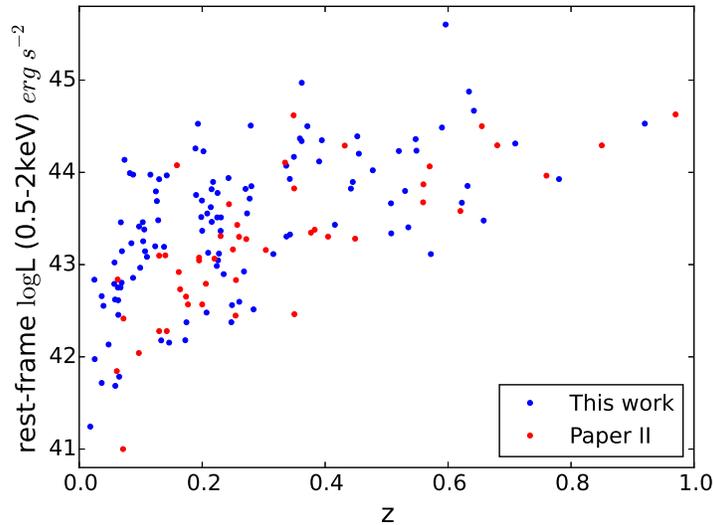}
\caption{Rest-frame 0.5-2 keV luminosity versus redshift.  More accurate luminosities measured with X-ray spectroscopy in Paper II are used when available (red points).}
\label{fig:L_z}
\end{figure}

If we consider also the X-ray redshift derived from the Telescopio Nazionale Galileo (TNG) observations and the X-ray spectral analysis of the sources in catalog I (Paper II) we have a total of \numb{158} sources with redshift, from optical spectroscopy or photometry, or from X-ray spectral analysis.  % 130+12+16=158
Therefore, about $60\%$ of our sample has redshift information.   For these sources, we plot the rest-frame 0.5-2 keV luminosities versus redshifts in Figure \ref{fig:L_z}.

\begin{figure*}[htbp]
\center
\includegraphics[width=0.32\columnwidth]{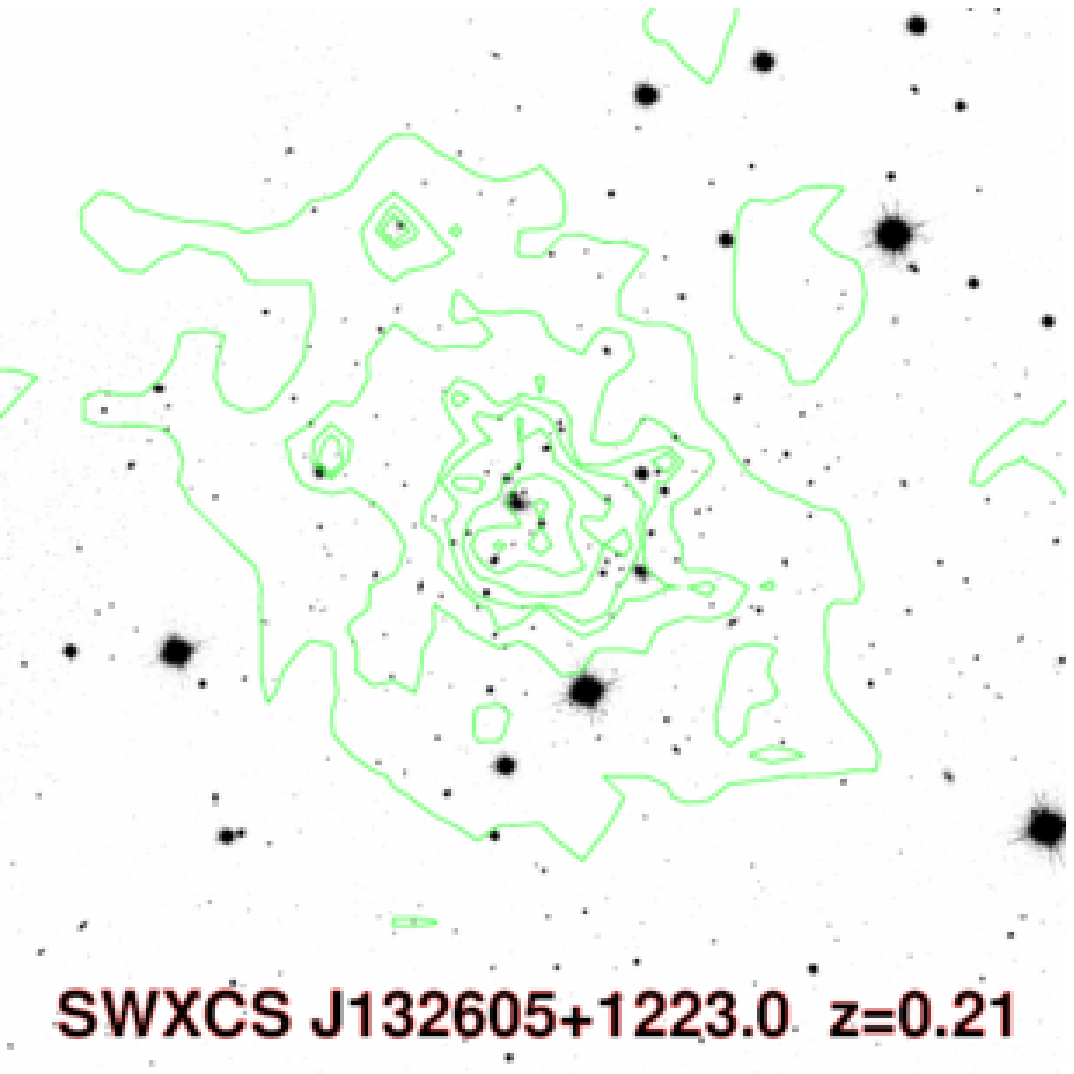}
\includegraphics[width=0.32\columnwidth]{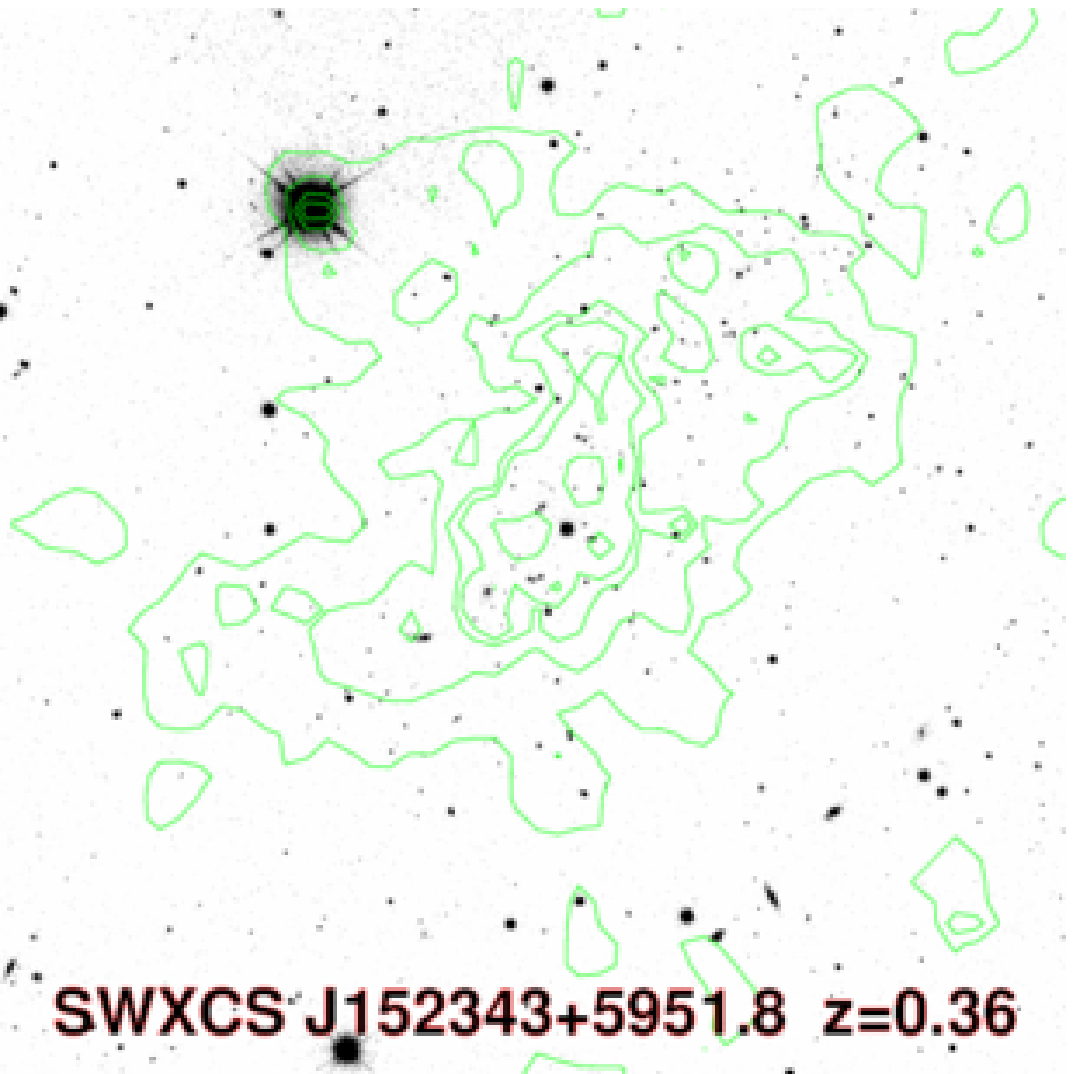}
\includegraphics[width=0.32\columnwidth]{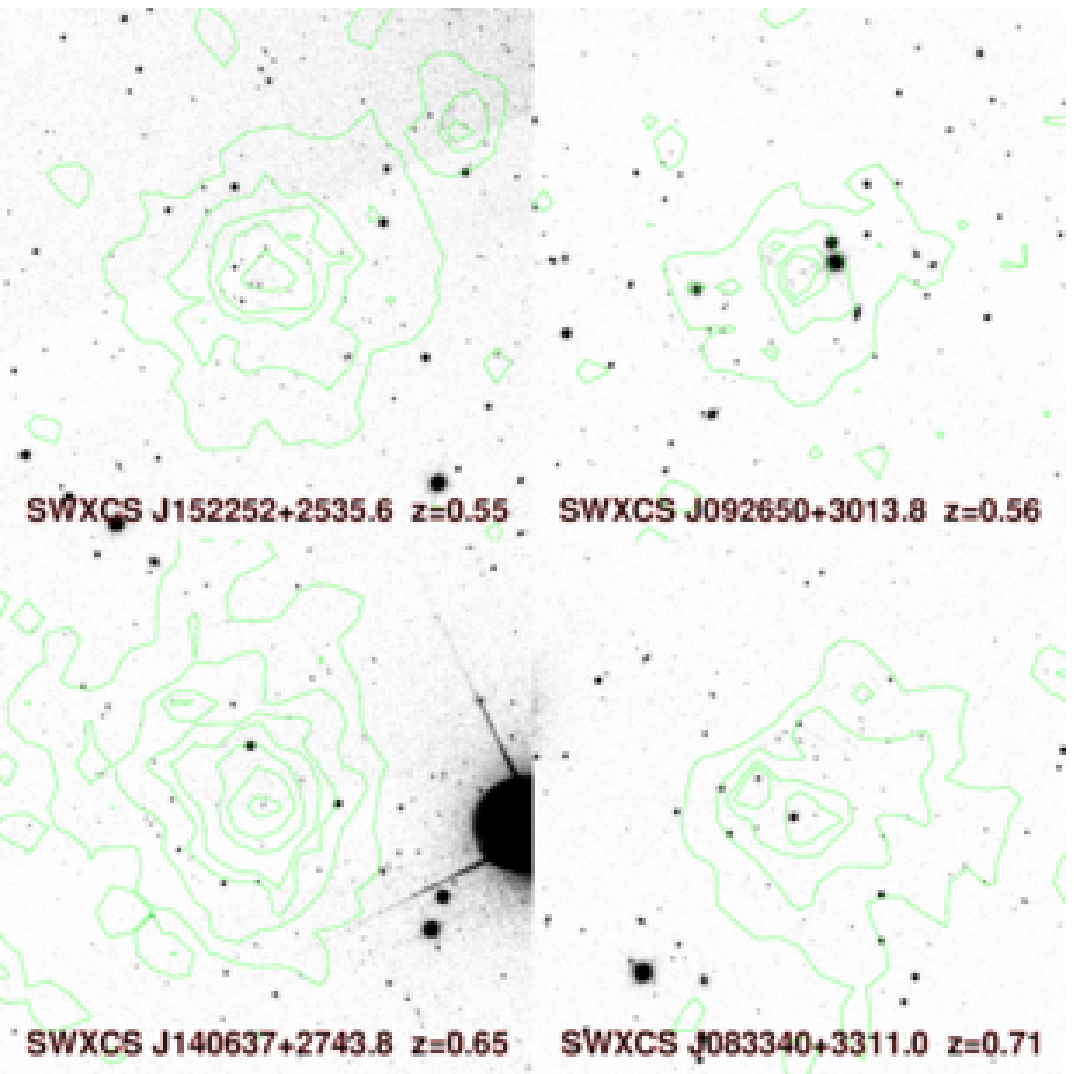}
\caption{
SDSS r-band images of a few medium-$z$ SWXCS sources with obvious optical counterparts.
The images have sizes of $5\arcmin\times5\arcmin$ or $10\arcmin\times10\arcmin$.
The X-ray contours (green lines) correspond to 2, 5, 10, 30, 70, 150, 300 times the local background.
}
\label{fig:sdssimages}
\end{figure*}

\begin{figure}[htbp]
\center
\includegraphics[width=0.29\columnwidth]{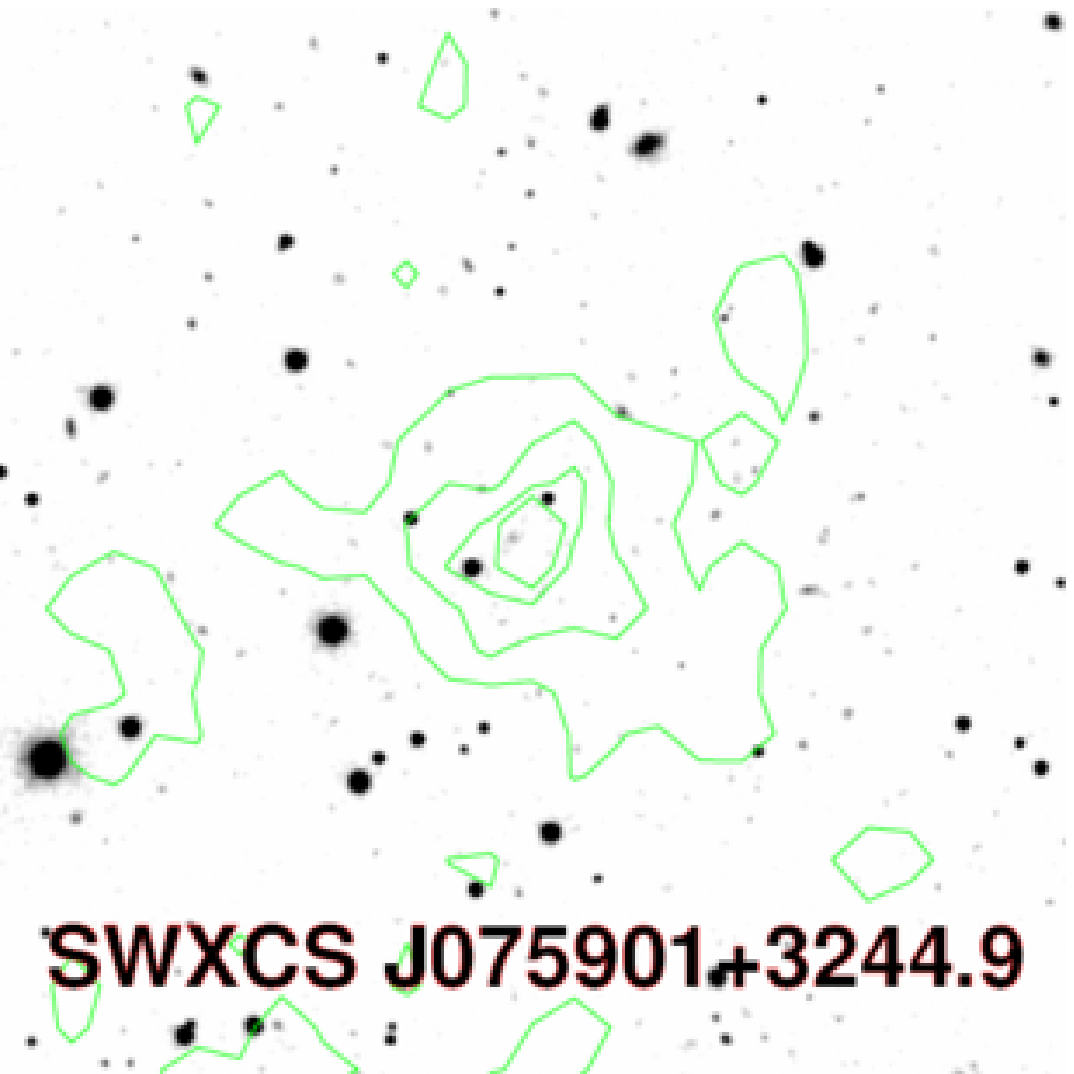}
\includegraphics[width=0.29\columnwidth]{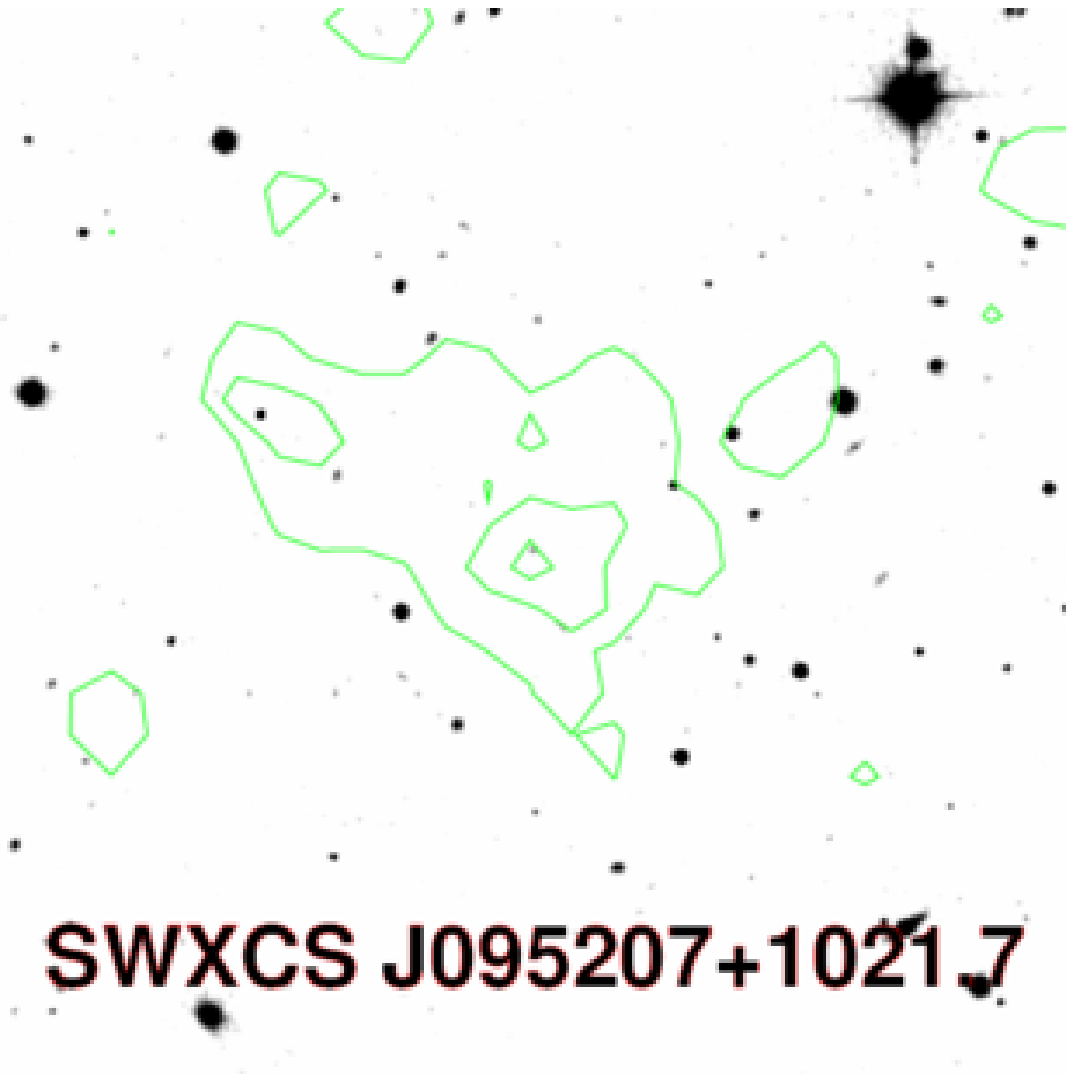}\\
\includegraphics[width=0.29\columnwidth]{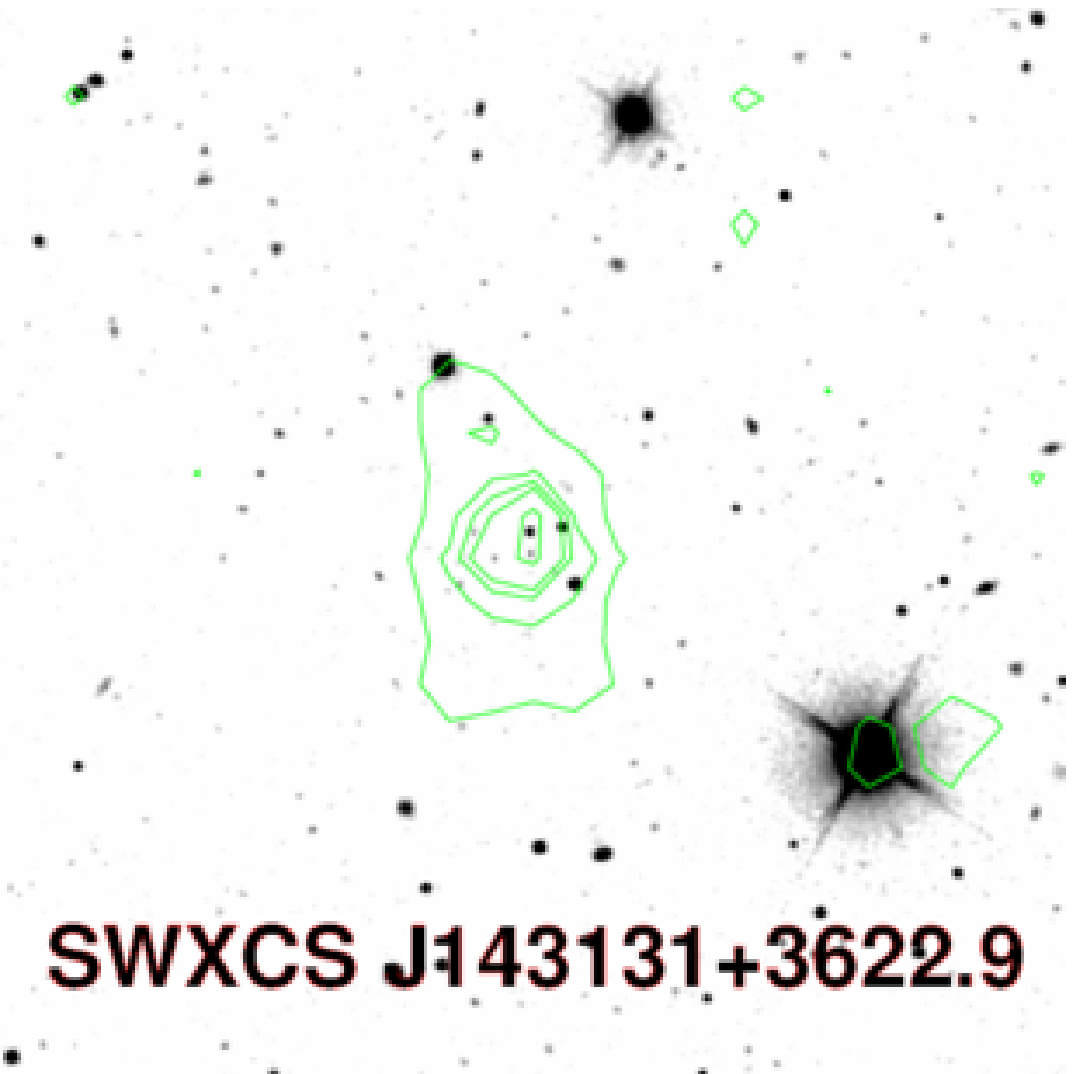}
\includegraphics[width=0.29\columnwidth]{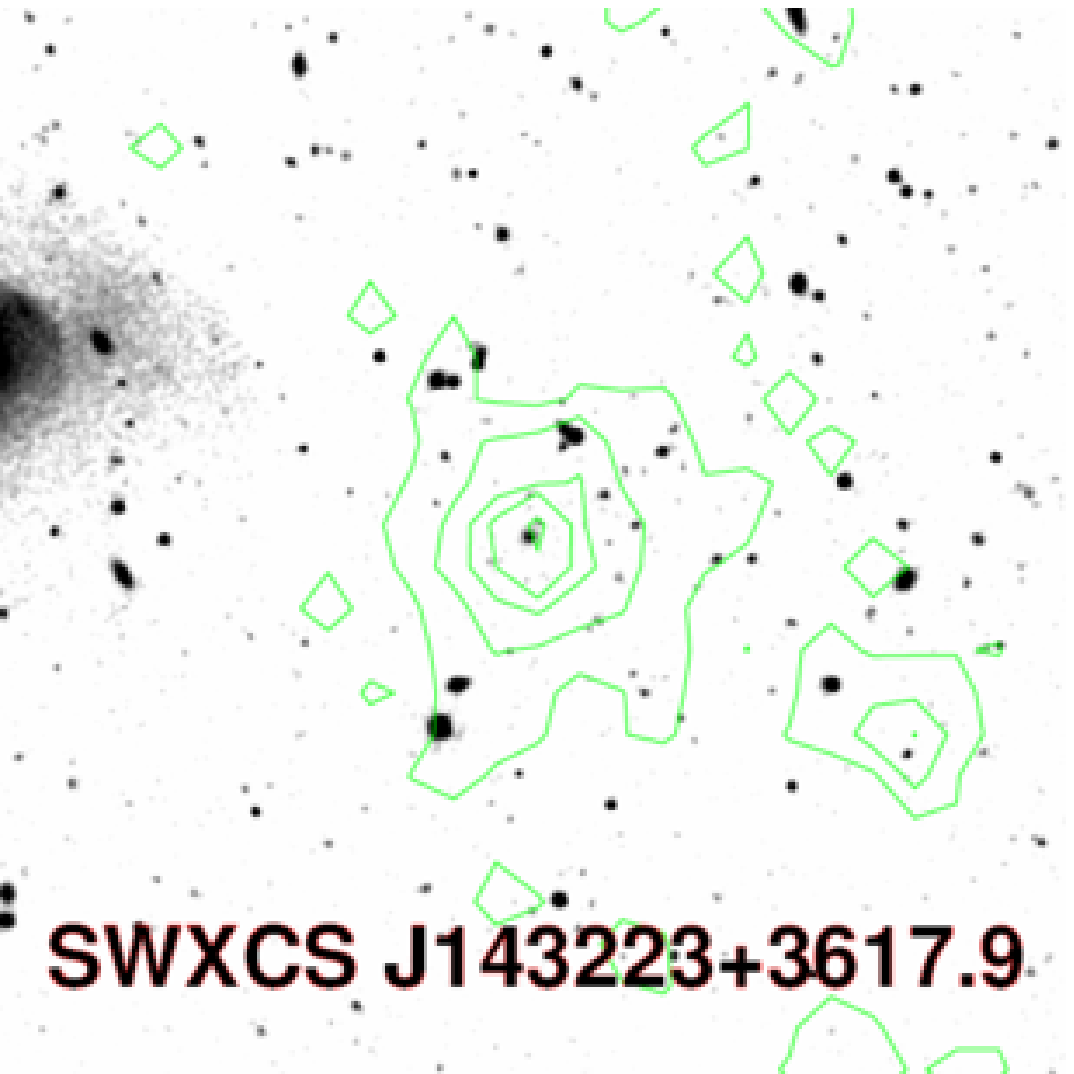}
\caption{SDSS r-band images of four SWXCS sources with soft-band fluxes less than $5\times10^{-14}$ \cgs and without obvious optical counterparts. 
These sources are among the high-$z$ cluster candidates in the SWXCS.
The image sizes are $5\arcmin\times5\arcmin$.
The X-ray contour generating method is the same as used in Figure \ref{fig:sdssimages}.
%The yellow contours show the \name{EXSdetect} extraction regions.
% An empty hole appears inside the region of SWXCS J000251.3-5258, which is due to fluctuation.
%SWXCS J000251.3-5258, SWXCS J062155.5-6228, SWXCS J094015.6-2153, SWXCS J215509.9+1650 whose soft band fluxes are 8, 7, 6, 6$\times10^{-15}$ \cgs.
}
\label{fig:high_z_candidates}
\end{figure}

We remark that \numb{116} sources overlap with SDSS images.  
In Figure \ref{fig:sdssimages} we show, as a sample, a selection of SDSS r-band images of SWXCS sources with obvious optical counterparts, with X-ray contours overlaid.
The X-ray and SDSS images (when available) for all the SWXCS sample can be found in the websites SWXCS website (\url{http://www.arcetri.astro.it/SWXCS/} and \url{http://swxcs.ustc.edu.cn}).
Among the source with optical redshift, the highest redshift is $z=0.92$ for XMMXCS J142908.4+424128.9 
\citep{Mehrtens12_XCS}.  This confirms that the depth of our catalog is sufficient to select clusters up to $z\sim 1$.  
Another high redshift source candidate is SWXCSJ011432-4828.4, whose redshift is  measured to be $0.97\pm0.02$ from X-ray spectral analysis in Paper II.  
The presence of clusters at $z\sim 1$ is not unexpected in the SWXCS, given the non negligible sky-coverage of the SWXCS at low fluxes.  
Based on previous results from the RDCS, we expect of the order of $\sim 10$ clusters with $z\geqslant 1$.
In addition to the few sources already mentioned, we already identified a sample of high-redshift candidates among the sources with SDSS images but no optical counterparts.
% The high-redshift candidates are naturally selected among the low-flux sources with no optical candidates in the SDSS images (when available).
As an example, the SDSS images of four of our high-$z$ cluster candidates are shown in Figure \ref{fig:high_z_candidates}.  

The next step of our project is to increase the number of identifications and redshift measurements, in order to use our sample for statistical studies and cosmological tests.  
We have started an extensive follow-up program to obtain sensitive, multi-band imaging photometry of the SWXCS sample.
Our immediate goal is to measure the integrated properties of the stellar populations of the galaxies through SED fitting.  We will also explore the correlation of the galaxy properties with those of 
the hosting cluster and their evolution with redshift.
The planned observations consist of deep CCD images in the UBVRIz filters which, in the redshift range $0.3<z<1$, probe the rest-frame spectral range from mid- and far-UV to the optical wavelengths.
%, i.e. from $2700\lesssim\lambda\lesssim 6500$ \AA\ at $z\sim 0.3$ to $1800\lesssim\lambda\lesssim 4500$ \AA\ at $z\sim 1$. Thus, throughout the redshift range spanned by the sample, the observations 
This choice allows us to properly sample the wavelengths across the 4000 \AA\ break, which are key for accurate measurement of the integrated stellar mass, star-formation rate, average dust obscuration 
and luminosity-weighted age through SED fitting.
For the southern sources we use the Du Pont 2.5-meter telescope at Las Campanas Observatory coupled with the Cassegrain-focus Direct Camera. 
%, equipped with a single thinned, back-illuminated, 2Kx2K CCD which provides a FOV 8.6x8.6 arcmin and a pixel scale of 0.259 arcsec/pix with excellent 
% image quality throughput throughout the UV, optical and near infrared window.
Sources in the northern emisphere are observed using the Mayall 4-meter telescope with the MOSAIC Prime Focus camera at the Kitt Peak National Observatory.
% This provides a much larger FOV of 36x36 arcmin with 8 abutting 2Kx4K CCD, also thinned and back-illuminated, with a plate scale of 0.26 arcsec/pixel.
%Typical exposure times are calibrated to secure at least SNR>10 in any band for a flat-spectrum source with r=24, and as much as SNR~40 in the g and r bands. 
% We estimate that this sensitivity will allow us to carry out SED fitting for galaxies at least down to $0.1L^*$ throughout the explored redshift range (the lower limit on mass varies with the 
% M/L of the galaxies; we should estimate it together with the minimum SFR we can measure). We have verified that these requirements allow us to get very good photometric redshifts and 
% stellar populations parameters on the clusters we have already reduced.
Currently, we have observed a total of 41 groups and clusters, 11 in the South and 30 in the North one.
We plan to release the reduced and calibrated images and source catalogs of the first year of observations  in early 2015. The program will continue in the following years. 

We finally note that we can add a significant number of redshifts by extending the X-ray 
spectral analysis to the entire sample. Although the requirements for a successful identification of the redshifted $K_\alpha$ Fe line, as shown in \citet{2011Yu} for {\sl Chandra}, do not apply to most of the SXCS sources, the lower background and the slightly better spectral resolution of \name{XRT} allows X-ray redshift measurements in a lower S/N regime, as shown in Paper II.  The X-ray spectral analysis of the SWXCS sample will be presented
in a forthcoming paper (Moretti et al. in preparation).

\section{CONCLUSIONS}
\label{conclusions}

We search for candidates groups and clusters of galaxies in $\sim 3000$ extragalactic 
\name{Swift-XRT} fields.  These fields are selected in order to provide a truly serendipitous survey, 
therefore excluding all the fields in the \name{Swift-XRT} archive which are somehow correlated with 
galaxy clusters and groups.  We use the software \name{EXSdetect}, which has been specifically
developed for this project and it is optimized for detection and photometry of extended sources 
in \name{Swift-XRT}  images.  Therefore, both in terms of covered solid angle and of sensitivity, this
work constitutes a significant extension of the first SWXCS catalog published in Paper I.

We find \numb{263} X-ray extended sources (including the \numb{71} sources already presented in Paper I) with negligible contamination and a well defined selection function.  The sky coverage ranges  from a maximum of  $\sim400$ deg$^2$ to 
$1$ deg$^2$ at a flux of $0.7\times10^{-14}$ \cgs.
The logN-logS is in very good agreement with previous deep surveys.
We cross-correlate SWXCS sources with previously published optical, X-ray or SZ cluster catalogs, finding that \numb{137} 
sources are already classified as clusters in any of the three bands, while \numb{126} sources are new cluster and group candidates.
We already collected redshifts measurement (both optical, spectroscopic of photometric, and X-ray)
for \numb{158} sources ($60$\% of the sample).  When the optical follow-up and the extension
of the X-ray spectral analysis will be completed, the SWXCS will provide a large and well defined
catalog of groups and clusters of galaxies to perform statistical studies of cluster properties
and tests of cosmological models.
All the results of the SWXCS are publicly available on http://www.arcetri.astro.it/SWXCS or http://swxcs.ustc.edu.cn, including machine-readable tables and the \name{EXSdetect} code.

\acknowledgements 
TL and JXW acknowledge support from National Basic Research Program of China (973 program, grant No. 2015CB857005) and Chinese National Science Foundation (grant No. 11233002, 11421303 and 11403021).
TL, AM, ET and PT received support from the "Exchange of Researchers" program for scientific and technological cooperation between Italy and the People's Republic of China for the years 2013-2015 (code CN13MO5).
We also thank the anonymous referee for helpful comments.
This research has made use of the NASA/IPAC Extragalactic Database (NED) which is operated by the Jet Propulsion Laboratory, California Institute of Technology, under contract with the National Aeronautics and Space Administration.  
% We also thank the anonymous Referee for comments and suggestions which 
% significantly improved the paper.

\section*{APPENDIX}
\renewcommand\thesubsection{\Alph{subsection}}
\subsection{Background Estimation\label{apd:bkgestimation}}

The background of an X-ray image is defined as the sum of all the recorded photons not associated to astronomical sources, or associated to some astronomical component that can not be resolved (like the Galactic diffuse emission).
Practically, we divide the photons in an X-ray image into two components: a background component with a roughly constant flux distributed randomly across the whole field, and an additional component, associated to single sources, with a highly concentrated spatial distribution covering only a very minor fraction of the field.
A well known result, obtained by numerical simulation \citep{Kiang66} is that for randomly positioned points, the distribution of the Voronoi cell areas follows an empirical formula:

\begin{equation}
\eqnum{A.1}
\label{kiang}
P(\tilde{f})\ =\ e^{-4/\tilde{f}}(\frac{32}{3\tilde{f}^3}+\frac{8}{\tilde{f}^2}+\frac{4}{\tilde{f}}+1)\, ,
\end{equation}
where $P$ is the cumulative probability distribution function (CDF, $P\in[0,1]$), $f = 1/a$ is the inverse of the cell area $a$, called flux here, and $\tilde{f}=f/\langle f\rangle$  is the flux normalized to the average value $\langle f\rangle=1/\langle a\rangle$.
Therefore, the distribution of the Voronoi cell areas is provided by a function of $f$ with only one parameter $\langle f\rangle$.
The value of $\langle f\rangle$ is the background of the image.
This relation was used in the first Voronoi algorithm for X-ray source detection proposed by \citet{Ebeling93}.
Photons in the faintest end of the filled-pixel distribution can be assumed to be only due to the background.
Therefore, they proposed an accurate measurement of the average background flux by fitting the faint-end CDF, specifically, only for the $f/\langle f\rangle<0.8$ part.
This part includes $\sim 27$\% of all the filled pixel in a pure-background image.  
We made use of this method in \name{EXSdetect} in \citet{Liu13_EXSdetect}.
Here we improved the background estimation running Monte Carlo simulations as described below.

We start from a direct test of Equation \ref{kiang} with simulations.
We randomly distribute one million photons in images of different sizes, chosen in order to have three different flux levels ($0.001, 0.03, 0.1$ photon/pixel).
Note that float-value positions are assigned to each photon, which is equivalent to assuming an infinitely small pixel scale for the images.
Using the \name{SweepLine} subtask of \name{EXSdetect}, we construct the Voronoi diagram for each image and calculate the area of each cell.
Then we compare Equation \ref{kiang} with the simulated distribution of fluxes.
We find that Equation \ref{kiang} describes the CDF well in the entire range, but not particularly in the faint end.
Therefore, the best fit of Equation \ref{kiang} for the $f/\langle f\rangle<0.8$ part does not recover the average background flux accurately. A systematic deviation is introduced.

Further more, the assumption of infinitely small pixel scale is not realistic.
Photon positions in real images are always in integer rather than in float.
In other words, digital images always have limited resolutions.
The limited resolution induces two effects: a distortion of the CDF in the faint-end which is due to the lower limit of Voronoi cell area, and a reduction of data points which is due to the fact that multiple photons on the same pixel contribute a single Voronoi cell.
These effects are flux dependent, being more significant at high fluxes.
In the simulation, we convert the photon positions into pixel positions, and test Equation \ref{kiang} in this realistic situation.
%The ratio of the measured background to the input background is shown in Table \ref{tab:bkgestimation} for nine combination of background count rates and exposure depths.
As shown in Table \ref{tab:bkgestimation}, the usage of Equation \ref{kiang} underestimate the average flux by 3-6\% in the flux range 0.001-0.1 photons/pixel.

The background measurement is improved as follows.
We add a parameter $c$ into Equation \ref{kiang}:
%, according to the original form of the formula used in \citet{Kiang66}, which contains a $\Gamma$ function.

\begin{equation}
\eqnum{A.2}
\label{modifiedkiang}
P(\tilde{f},c)\ =\ e^{-c/\tilde{f}}(\frac{c^3}{6\tilde{f}^3}+\frac{c^2}{2\tilde{f}^2}+\frac{c}{\tilde{f}}+1) \, .
\end{equation}

\begin{figure}[htbp]
\figurenum{A.1}
\center
\plotone{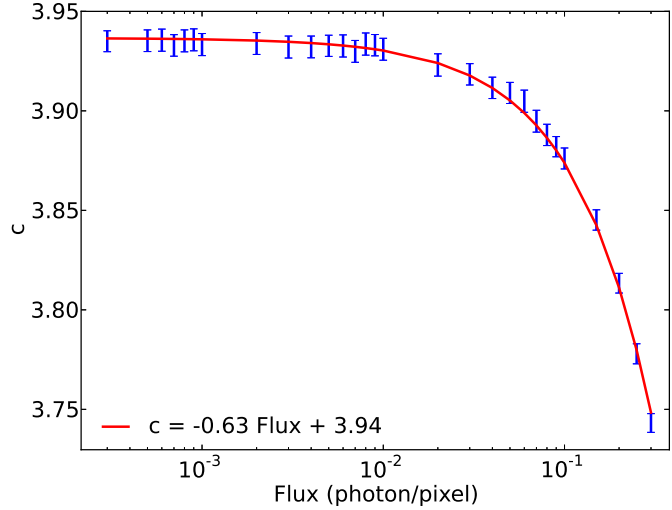}
\caption{The correlation between the best fit $c$ parameter in Equation \ref{modifiedkiang},
 and the input background. The red line is the best fit Equation \ref{bestfitc}.}
\label{fig:c_flux}
\end{figure}

Equation \ref{modifiedkiang} corresponds to Equation \ref{kiang}  when $c=4$.
Then we repeat the fit of the simulated CDF in the region $f/\langle f\rangle < 0.8$ and search for value of the $c$ parameter which minimize the difference between the model and the simulated data.  We also replace the definition of average flux 
$\langle f\rangle=1/\langle a\rangle=N_{pixels}/\sum{a}$ with 
$\langle f\rangle=N_{photons}/\sum{a}$, 
to take into account multiple photons in the same pixel.
As shown in Figure \ref{fig:c_flux}, we find a linear correlation between $c$ and $\langle f\rangle$:

\begin{equation}
\eqnum{A.3}
\label{bestfitc}
c\ =\ -0.63\langle f\rangle\ + 3.94 \, .
\end{equation}

Equation \ref{modifiedkiang} and \ref{bestfitc} allow us to recover the background accurately with no deviation, as shown in Table \ref{tab:bkgestimation}.

% Thus, we have two improved formulae (TWO?) to describe the CDF of random photons. 
% They still serve as a function of $f$ with only one parameter $\langle f\rangle$, 
% which fits the $\langle f\rangle<0.8$ part of the CDF at any flux level.
% Using these improved formulae, we can measure the average background flux 
% for any background regime,  see
% see the illustration in the lower panel of Figure \ref{fig:diffbkg} and 
% the quantitative estimations in Table \ref{tab:bkgestimation}.

%\begin{figure}[htbp]
%\center
%\includegraphics[width=0.7\columnwidth]{figures/diffbkg_float.eps}
%\includegraphics[width=0.7\columnwidth]{figures/diffbkg_Kiang66.eps}
%\includegraphics[width=0.7\columnwidth]{figures/diffbkg_improved.eps}
%\caption{For simulated images with one million photons distributed randomly inside, at three 
%different flux levels ($0.001, 0.03, 0.1\ photon/pixel$), we measure their CDFs of normalized inverse cell area $f/\langle f\rangle$, and plot the residuals ($\Delta P(f/\langle f\rangle)$) data 
%minus model (red lines).
%For the blue lines, best-fit model is used instead of the model with the known input 
%$\langle f\rangle$.  Green boxes stress the $\langle f\rangle<0.8$ range.
%The raw photon position (float value) lists are used in the upper panel, and they are converted 
%into pixel-by-pixel images (with integer-value photon positions) in the middle and lower panel.
%The \citet{Kiang66} formula is used in the upper and middle panels, and our improved model is 
%shown in the lower one.
%}
%\label{fig:diffbkg}
%\end{figure}

\begin{table}[htbp]
\tablenum{A.1}
\begin{center}
\caption{Background estimation accuracy, measured by the ratio (percentage) of output flux to 
input flux in simulation. Three levels of background flux (in unit of $photon/pixel$) and three cases 
of exposure depth (different number of photons within one image) are considered.}
\label{tab:bkgestimation}
\begin{tabular}{l|ccc}
\hline
Flux	&N=100	&N=1000	&N=100000\\
\hline
&\multicolumn{3}{c}{\citet{Kiang66} formula}\\
\hline
0.001	&96.5$\pm$6.3	&97.4$\pm$2.2	&97.4$\pm$0.2\\
0.03	&95.9$\pm$6.4	&96.8$\pm$2.2	&96.9$\pm$0.2\\
0.1	&94.2$\pm$6.1	&95.5$\pm$2.0	&95.7$\pm$0.2\\
\hline
&\multicolumn{3}{c}{improved formulae}\\
\hline
0.001	&98.6$\pm$6.8	&99.7$\pm$2.3	&99.8$\pm$0.2\\
0.03	&98.7$\pm$7.0	&99.8$\pm$2.3	&100.0$\pm$0.2\\
0.1	&98.6$\pm$6.8	&100.0$\pm$2.2	&100.1$\pm$0.2\\
\hline
\end{tabular}
\end{center}
\end{table}

A main limitation of our background measurement method consists in the background variation across the image due to vignetting.
In order to take vignetting into account, in \citet{Liu13_EXSdetect}, the background map of the field, in units of $photon/pixel$, was obtained by multiplying the average background flux by the vignetted exposure map.
However, only the background from astronomical sources is vignetted, the background components associated to instrumental noise and cosmic rays vary following a more constant pattern.
Therefore, the simple procedure adopted in \citet{Liu13_EXSdetect} over-corrected the vignetting effect, especially at the image borders.

\begin{figure}[htbp]
\epsscale{0.5}
\figurenum{A.2}
\center
\plotone{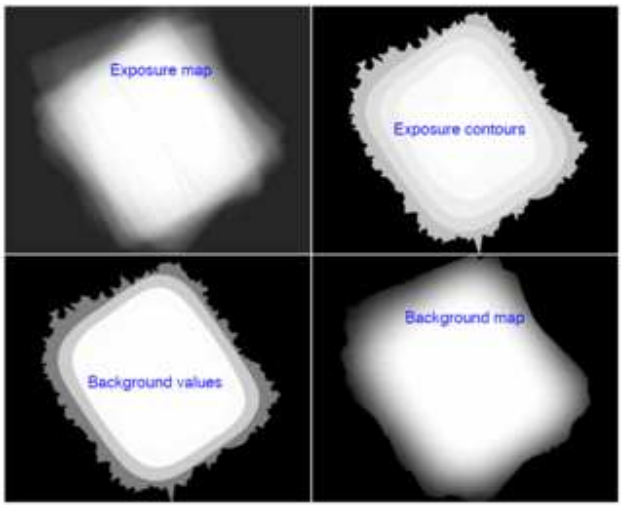}
\caption{{\sl Top left:} exposure map of a SWXCS field; {\sl top right:} concentric regions
are obtained according to the exposure map values; {\sl bottom left:} the background is computed
in each concentric region by applying Equation \ref{modifiedkiang}; {\sl bottom right:} a continuous
background image is obtained.}
\label{fig:bkgsteps}
\end{figure}

In the new version of \name{EXSdetect} used in this work, we refined the background estimation in the following way.
First we divide the field into about five concentric regions delimited by the smoothed contours on the exposure map (see upper panels of Figure \ref{fig:bkgsteps}). 
We calculate the average background flux in each of these regions with the method based on the improved \citet{Kiang66} formula, creating a step-like background map.
Using the background fluxes and exposure times in these bins, we interpolate the relation between the background and the exposure time values in each concentric region with a linear regression.
Applying this relation to the original exposure map directly provides a continuous background map 
(lower panels of  Figure \ref{fig:bkgsteps}).
The background map is used to recover the background value at the source position when applying the criterion to define source region.

\subsection{Source Classification\label{apd:classification}}

High angular resolution is a crucial parameter for detection 
of extended sources in X-ray images.  With HEW of the order of $5$\arcsec, 
it is possible, in principle, to identify any extended source associated to clusters and groups
up to high redshift. The only critical aspect left in high resolution X-ray images concerns
the discrimination of extended features with very low surface brightness 
from background fluctuations.  In Wolter Type I X-ray mirrors, the angular resolution is 
maximized at the aimpoint, while at large off-axis angle the PSF rapidly degrades.
This aspect creates several problems when searching for serendipitous extended sources 
across an image, particularly if the PSF is not axisymmetric and if the image is obtained by the
merging of many exposures.

\name{Swift-XRT} has the valuable property, unique among existing X-ray facilities, 
of showing a constant PSF across the FOV, but the price 
to pay is a moderate angular resolution.  In many cases, the extent of the image of an unresolved
source is not very far from the extent of a genuine, compact extended source, like
clusters at high-$z$, cool core small clusters, etc.  To minimize this effect, we significantly
improved the source classification method of \name{EXSdetect} with respect to the version of
\citet{Liu13_EXSdetect}.

\begin{figure}[htbp]
\epsscale{0.5}
\figurenum{B.1}
\center
\plotone{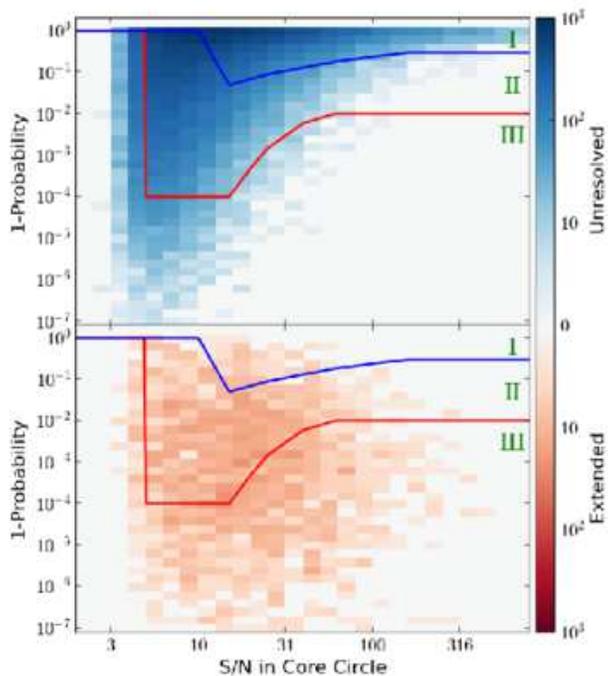}
\caption{Source classification curves in the $1-P$-S/N space, where $P$ is the 
probability of a given source to be extended, based on a comparison between the
surface brightness distribution and a PSF model within $5$ pixels 
(see \citet{Liu13_EXSdetect} for details).   The 
selection criteria (blue and red lines) classify the sources into three categories: I =  
unresolved; II= ambiguous; III= extended sources.
The color-coded grid shows the number of simulated source in the $1-P$-S/N space, 
separately for unresolved (upper panel) and extended (lower panel) sources.}
\label{fig:boundary}
\end{figure}

In \citet{Liu13_EXSdetect} the source image and the PSF model were compared inside a 
circle with a radius of 5 pixels (corresponding to a $\sim60\%$ encircled energy at 1.5 keV)
to take advantage of the larger S/N ratio with respect to the outer part of the PSF.
In other words, the difference in the profile of the extended source image with respect 
to the PSF model within 5\arcsec was sufficient to identify it as extended.
Sources were then divided into three types: I unresolved sources; 
II ambiguous sources; III extended sources. These selection thresholds were obtained based
on our simulations, as shown in Figure \ref{fig:boundary}.  In this Figure we show the
value $1-P$ where $P$ is the probability of a source of being extended, corresponding to 
our selection threshold  as a function of the S/N.  Color coded is the density, in the $1-P$-$S/N$
space, of the simulated unresolved (upper panel of Figure \ref{fig:boundary}) and of the 
simulated extended (lower panel of Figure \ref{fig:boundary}) sources.

In the most recent version of \name{EXSdetect}, we introduce a further step to classify the sources
lying between the unresolved and extended regions.  We argue that, in addition to the
profile of the surface brightness distribution, an additional information is contained
in the shape of a source.  In particular, any unresolved source is expected to have 
approximately circularly symmetric isocontours, according to the PSF model.
To recover this information we consider a larger radius of 7 pixels
(corresponding to $\sim70\%$ encircled energy for an unresolved source).  The PSF model
clearly provides the flux level of an unresolved source at a radius of 7 pixels.  Then, 
a FOF algorithm is run on all the pixels whose flux
is larger than the threshold.  If the source is truly unresolved, this
region should be very close to a circle with a radius of 7 pixels.
Significant emission detected outside this radius, is taken as a hint of an extended source.
On the other hand, if several pixels fall below this value within the circle of 7 pixels, 
the source is most likely unresolved.
% THIS IS DONE BY EYE - MUST BE MORE QUANTITATIVE HERE

\begin{figure}[htbp]
\figurenum{B.2}
\center
\includegraphics[width=0.28\columnwidth]{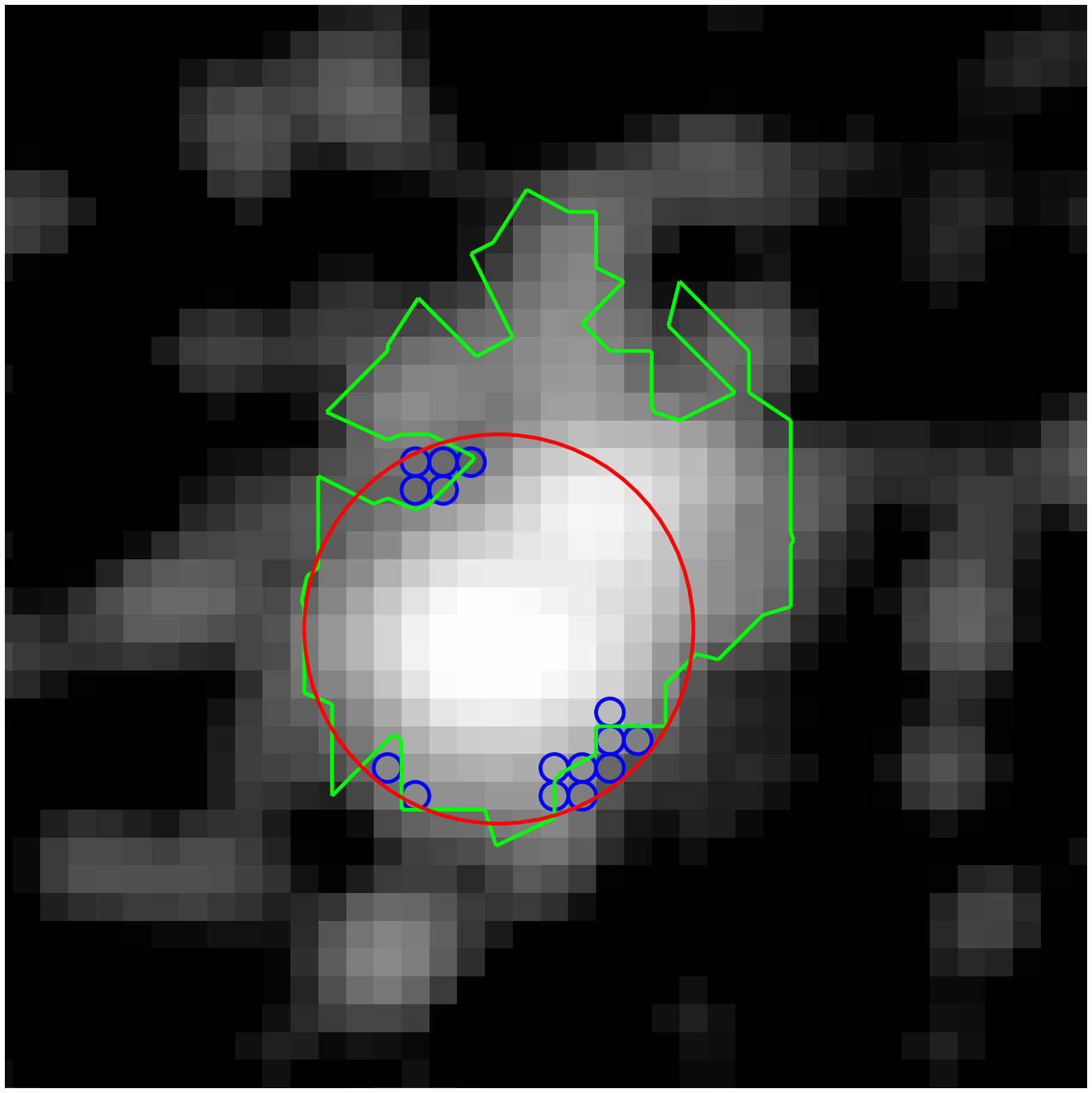}
\includegraphics[width=0.28\columnwidth]{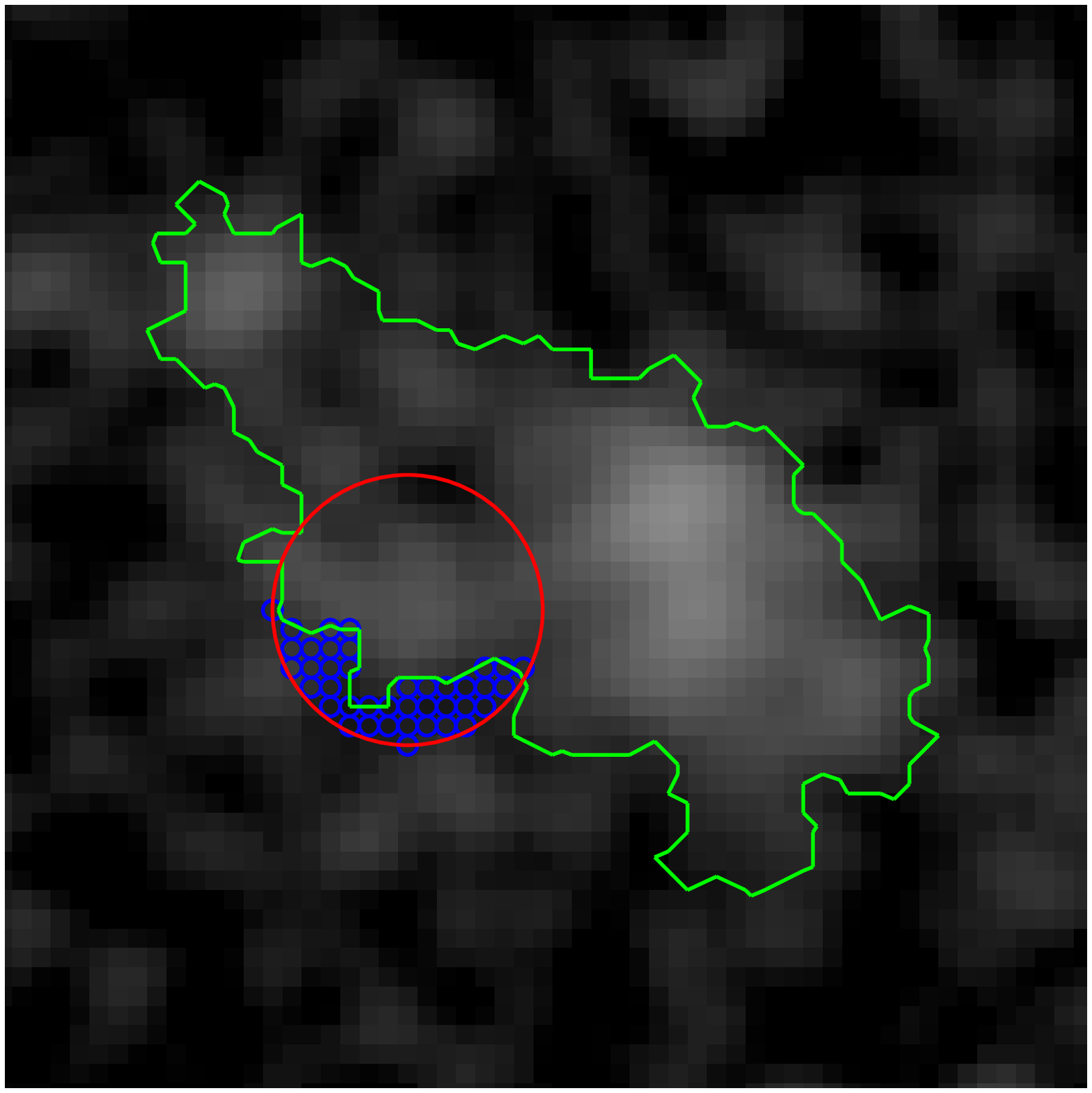}\\
\includegraphics[width=0.28\columnwidth]{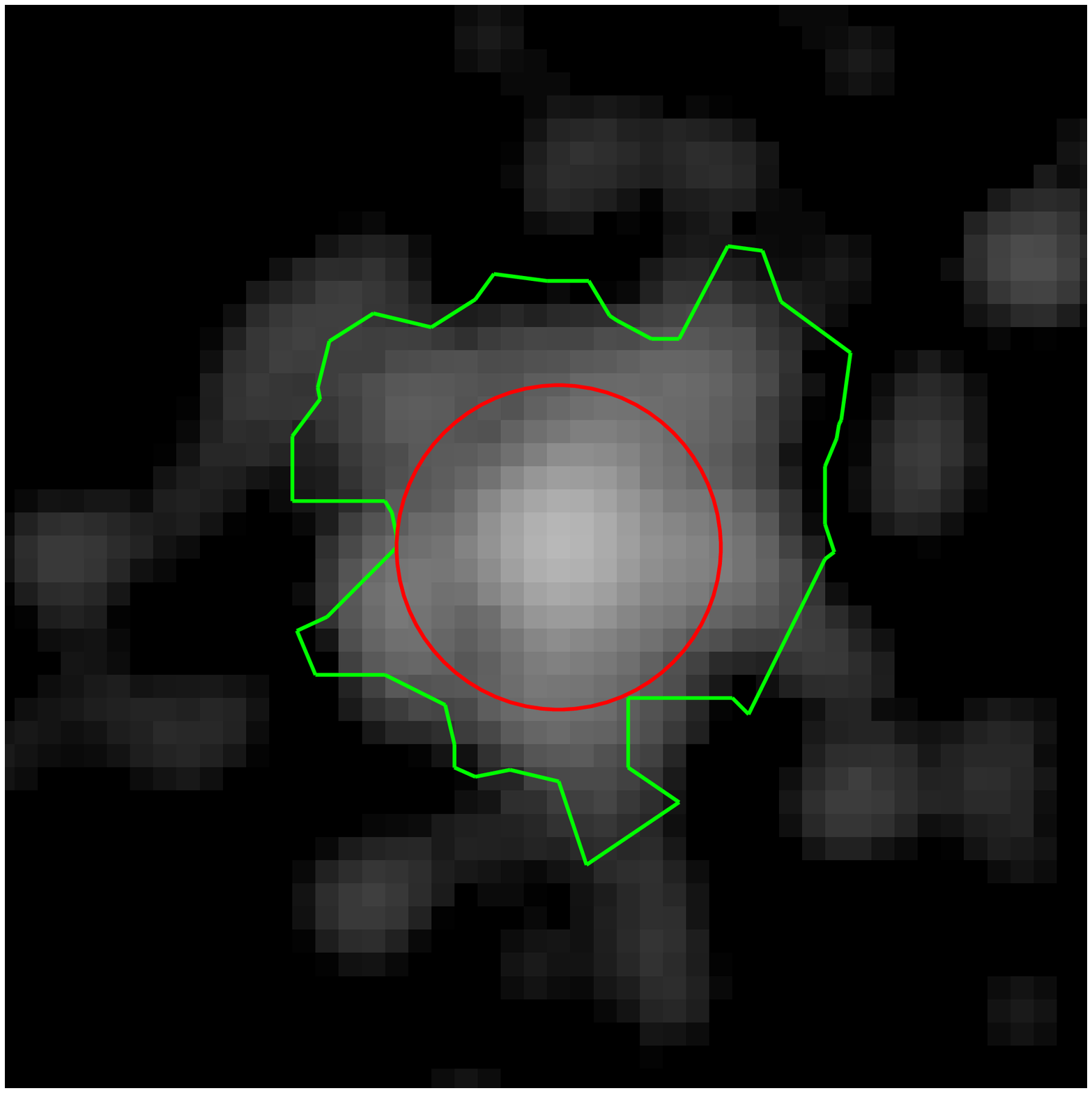}
\includegraphics[width=0.28\columnwidth]{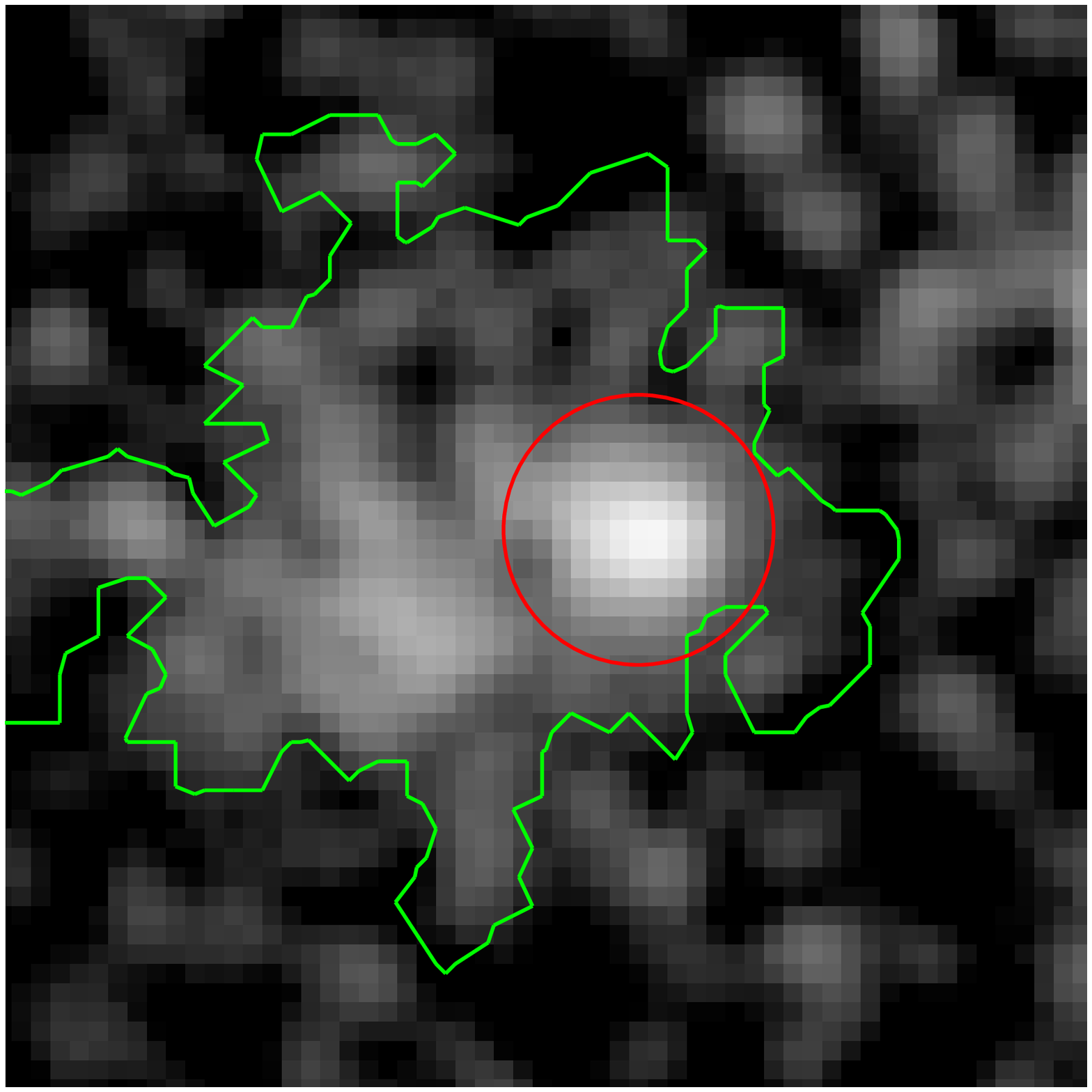}
\caption{Four examples of disambiguations in real \name{XRT} images.  Red circles
mark the position of the candidate extended source, and have a radius of 7 pixels.
Green lines enclose the FOF regions which reach the same flux level as the PSF 
at the radius of 7 pixels.  Blue circles mark pixels outside the FOF regions 
but inside the red circles.  The two sources in the upper panels
% (23:33:32.1,-66:17:07 06:22:00.2,-62:28:46) 
are classified as unresolved, while the sources in the lower panels 
% (10:20:32.3,-02:28:25 and 11:45:49.3,+59:53:25) 
are classified as extended.
}
\label{fig:srcclass7}
\end{figure}

A few cases of source disambiguation are shown in Figure \ref{fig:srcclass7}.
Unresolved sources can be identified with this method even under the contamination 
of other nearby sources, both brighter and fainter 
(upper panels of Figure \ref{fig:srcclass7}).  On the other hand, extended sources 
may by misclassified for several reasons: some may show a very low S/N in the core;
some may harbor very compact  cores; some others simply harbor a bright unresolved source embedded in the diffuse emission.  All these
cases can be identified simply applying our disambiguation criterion 
(see lower panels of Figure \ref{fig:srcclass7}). 
We tested against simulations that this criterion is efficient when the extended emission 
is above the background across a region with an effective radius $R_{eff} \sim 33$\arcsec.  
Clearly, the angular resolution of the instrument constitute a hard limit below which extended sources
can not be identified by any mean. In the case of SWXCS, all the extended sources 
with $R_{eff} $ close to the hard limit set by the $HEW = 18$\arcsec can not be identified
as extended.  Indeed, the minimum size of the sources in the SWXCS 
corresponds to $R_{eff} = 27$\arcsec.  

\end{CJK*}
\bibliography{sxcs}
\clearpage
\begin{deluxetable}{lrrcrrllr}
\centering
\tablenum{2}
\tablewidth{0pt}
\tabletypesize{\footnotesize}
\tablecaption{\label{tab:catalog} 
SWXCS catalog. Sources marked with an asterisk are included in the first relaease of SWXCS (Paper I; Paper II).
Column 1: source name according to the IAU Registry; sources included in the first release keep the name used in Paper II despite the new centroid positions.
Column 2-3: RA and Dec (J2000) coordinates of the X-ray centroid in degree.
Column 4: effective exposure time at the source position in sec.
Column 5: Galactic $HI$ columns density in $10^{20}$ cm$^{-2}$.
Column 6: effective radius $R_{eff}$ in arcsec; the area of the source region is $\pi R_{eff}^2$.
Column 7: net counts in the source region in the 0.5-2 keV band with 1 $\sigma$ error.
Column 8: S/N in the 0.5-2 keV band.
Column 9: flux in units of $10^{-14}$ \cgs in the 0.5-2 keV band, with 1 $\sigma$ error.
}
\tablehead{
\colhead{Name}	&\colhead{Ra}	&\colhead{Dec}	&\colhead{$t_{eff}$}	&\colhead{$N_H$}	&\colhead{$R_{eff}$}	&\colhead{$N_{net}$}	&\colhead{SNR}	&\colhead{Flux}
}
\startdata
SWXCS J000251-5258.5	&0.713858	&-52.974476	&304043	&1.59	&39.3	&104$\pm$17	&6.3	&0.8$\pm$0.1\\
SWXCS J000315-5255.2*	&0.813067	&-52.915205	&313159	&1.59	&79.8	&1089$\pm$43	&25.5	&8.5$\pm$0.5\\
SWXCS J000324-5253.8*	&0.846572	&-52.899251	&294650	&1.59	&90.5	&1012$\pm$43	&23.3	&8.4$\pm$0.5\\
SWXCS J000345-5301.8*	&0.934069	&-53.030934	&308065	&1.61	&67.5	&362$\pm$30	&12.2	&2.9$\pm$0.3\\
SWXCS J002044-2544.0	&5.181310	&-25.733882	&3272	&2.43	&217.2	&237$\pm$19	&12.5	&180.8$\pm$16.2\\
SWXCS J002114+2059.7	&5.309518	&20.995620	&95736	&3.86	&69.9	&90$\pm$14	&6.4	&2.4$\pm$0.4\\
SWXCS J002437-5803.9*	&6.157687	&-58.064728	&71839	&1.22	&114.4	&360$\pm$24	&14.7	&12.2$\pm$1.0\\
SWXCS J002729-2326.1	&6.870954	&-23.435391	&28798	&1.75	&99.2	&104$\pm$13	&7.9	&8.9$\pm$1.2\\
SWXCS J002824+0927.1	&7.098729	&9.451710	&44478	&3.95	&99.8	&121$\pm$15	&7.9	&7.1$\pm$0.9\\
SWXCS J002826+0918.3	&7.110244	&9.304376	&45409	&3.76	&102.4	&140$\pm$16	&8.6	&8.0$\pm$1.0\\
SWXCS J003316+1939.4*	&8.319273	&19.656753	&44727	&3.98	&76.1	&93$\pm$13	&7.3	&5.4$\pm$0.8\\
SWXCS J003759-2504.4	&9.494119	&-25.073476	&7612	&1.45	&217.7	&94$\pm$14	&6.8	&30.1$\pm$4.6\\
SWXCS J004311-1129.3	&10.796565	&-11.488947	&9586	&2.38	&100.2	&87$\pm$11	&8.3	&22.7$\pm$2.9\\
SWXCS J005059-0929.5	&12.744153	&-9.491900	&30035	&3.51	&131.6	&255$\pm$20	&12.8	&21.9$\pm$1.9\\
SWXCS J005500-3852.4*	&13.750356	&-38.874722	&40510	&3.31	&106.2	&136$\pm$17	&8.1	&8.6$\pm$1.1\\
SWXCS J010030-4749.4	&15.126057	&-47.823462	&15702	&1.89	&129.2	&112$\pm$14	&8.2	&17.7$\pm$2.3\\
SWXCS J010955-4555.9	&17.481203	&-45.930912	&198459	&1.92	&484.5	&40091$\pm$257	&156.1	&498.5$\pm$20.2
\enddata
\tablecomments{This table is available in its entirety in machine-readable form.}
\end{deluxetable}

\clearpage
\begin{deluxetable}{llllcccr}
\rotate
\centering
\tablenum{3}
\tabletypesize{\scriptsize}
\tablewidth{0pt}
\tablecaption{\label{tab:matched} Catalog cross-correlation results.
(1) Source name;
(2) optical redshift of cluster or galaxy counterparts, (p) for photometric;
(3) TNG measured redshift from Paper II;
(4) X-ray redshift from Paper II;
(5) catalog where the cluster counterpart is from;
(6) cluster counterparts;
(7) galaxy counterparts within 7 arcsec found in NED;
(8) separation of the matches in arcmin.
}
\tablehead{
\colhead{Name}	&\colhead{$z_{opt}$}	&\colhead{$z_{TNG}$}	&\colhead{$z_{X}$}	&\colhead{Catalog}	&\colhead{Cluster counterpart}	&\colhead{Galaxy counterpart}	&\colhead{Separation}}
\startdata
SWXCS J000315-5255.2*	&	&	&$0.62\pm0.1$	&	&	&	&0.00\\
\tableline
SWXCS J000324-5253.8*	&	&	&$0.76\pm0.01$	&	&	&	&0.00\\
\tableline
SWXCS J002044-2544.0	&0.141	&	&	&Planck	&Planck 119	&	&1.22\\
	&0.1410	&	&	&REFLEX	&RXC J0020.7-2542	&	&1.45\\
	&0.1410	&	&	&MCXC	&MCXC J0020.7-2542	&	&1.46\\
	&0.142352	&	&	&Abell	&Abell 0022	&	&2.08\\
\tableline
SWXCS J002437-5803.9*	&	&	&$0.195\pm0.012$	&	&	&	&0.00\\
\tableline
SWXCS J002824+0927.1	&0.2238(p)	&	&	&WHL	&WHL J002827.3+092612	&	&1.26\\
\tableline
SWXCS J002826+0918.3	&0.2258(p)	&	&	&WHL	&WHL J002826.9+091824	&	&0.17\\
\tableline
SWXCS J003759-2504.4	&0.063600	&	&	&Abell	&Abell 2800	&	&0.68\\
\tableline
SWXCS J005059-0929.5	&0.2000	&	&	&WHL	&WHL J005058.1-092923	&	&0.19\\
	&0.19997	&	&	&GMBCG	&GMBCG J012.74197-09.48965	&	&0.19\\
	&0.200058	&	&	&MaxBCG	&MaxBCG J012.74197-09.48965	&	&0.19\\
	&0.199	&	&	&400d	&400d J0050-0929	&	&0.33\\
	&0.1990	&	&	&MCXC	&MCXC J0050.9-0929	&	&0.34\\
\tableline
SWXCS J005500-3852.4*	&	&	&	&EDCC	&EDCC 493	&	&0.96\\
	&  0.164127	&	&	&NED	&	&LCRS B005239.6-390844         	&0.11\\
\tableline
SWXCS J010955-4555.9	&0.0238	&	&	&MCXC	&MCXC J0110.0-4555	&	&0.84\\
	&0.0238	&	&	&REFLEX	&RXC J0110.0-4555	&	&0.87\\
	&0.024700	&	&	&Abell	&Abell 2877	&	&2.44\\
	&0.0238	&	&	&Planck	&Planck 1024	&	&3.29\\
\tableline
SWXCS J011432-4828.4*	&	&	&$0.97\pm0.02$	&	&	&	&0.00\\
\tableline
SWXCS J012457-8104.9	&0.063377(p)	&	&	&Abell	&Abell S0158	&	&0.13\\
	&  0.087637	&	&	&NED	&	&2MASX J01245648-8104579       	&0.03\\
\enddata
\tablecomments{This table is available in its entirety in machine-readable form.}
\end{deluxetable}
%\clearpage
%\end{landscape}

\end{document}